\documentclass[aps,prc,preprint,amsmath,amssymb,showpacs,superscriptaddress]{revtex4}

\usepackage{graphicx}% Include figure files
\usepackage{dcolumn}% Align table columns on decimal point
\usepackage{bm}% bold math
\usepackage{longtable}
\usepackage{color}
\usepackage{CJK}

\begin{document}

\title{Covariant density functional theory for antimagnetic rotation}

\author{P. W. Zhao}%
\affiliation{State Key Laboratory of Nuclear Physics and Technology, School of Physics, Peking University, Beijing 100871, China}
\author{J. Peng }%
\affiliation{Department of Physics, Beijing Normal University, Beijing 100875, China}
\author{H. Z. Liang}%
\affiliation{State Key Laboratory of Nuclear Physics and Technology, School of Physics, Peking University, Beijing 100871, China}
\author{P. Ring}%
\affiliation{State Key Laboratory of Nuclear Physics and Technology, School of Physics, Peking University, Beijing 100871, China}
\affiliation{Physik Department, Technische Universit\"at M\"unchen, D-85748 Garching, Germany}
\author{J. Meng}%
\affiliation{State Key Laboratory of Nuclear Physics and Technology, School of Physics, Peking University, Beijing 100871, China}
\affiliation{School of Physics and Nuclear Energy Engineering, Beihang University, Beijing 100191, China}
\affiliation{Department of Physics, University of Stellenbosch, Stellenbosch, South Africa}

\date{\today}
\begin{abstract}
Following the previous letter on the first microscopic description of the antimagnetic rotation (AMR) in $^{105}\rm Cd$, a systematic investigation and detailed analysis for the AMR band in the framework of tilted axis cranking (TAC) model based on covariant density functional theory are carried out. After performing the microscopic and self-consistent TAC calculations with an given density functional, the configuration for the observed AMR band in $^{105}\rm Cd$ is obtained from the single-particle Routhians. With the configuration thus obtained, the tilt angle $\theta_\Omega$ for a given rotational frequency is determined self-consistently by minimizing the total Routhian with respect to the angle $\theta_\Omega$. In such a way, the energy spectrum, total angular momenta, kinetic and dynamic moments of inertia, and the $B(E2)$ values for the AMR band in $^{105}\rm Cd$ are calculated. Good agreement with the data is found. By investigating microscopically the contributions from neutrons and protons to the total angular momentum, the ``two-shears-like'' mechanism in the AMR band is clearly illustrated. Finally, the currents leading to time-odd mean fields in the Dirac equation are presented and discussed in detail. It is found that they are essentially determined by the valence particles and/or holes. Their spatial distribution and size depend on the specific single-particle orbitals and the rotational frequency.
\end{abstract}

% 21.60.Jz Nuclear Density Functional Theory and extensions (
%includes Hartree¨CFock and random-phase approximations)
%21.10.-k Properties of nuclei; nuclear energy levels
%21.10.Re Collective levels
%21.60.Ev Collective models
%23.20.-g Electromagnetic transitions
%23.20.Js Multipole matrix element
%27.60.+j 90 ¡Ü A ¡Ü 149
%27.50.+e 59 ¡Ü A ¡Ü 89

\maketitle

\section{Introduction}

In the previous letter~\cite{Zhao2011Phys.Rev.Lett.122501}, we reported the first fully self-consistent and microscopic investigation of antimagnetic rotation (AMR) in $^{105}\rm Cd$ in the framework of the tilted axis cranking (TAC) model based on the covariant density functional theory (CDFT). We found that the experimental data of this band~\cite{Choudhury2010Phys.Rev.C61308} have been reproduced rather well without any adjustable parameters. The present paper goes in more details along this direction by providing a series of interesting results for the AMR band in $^{105}\rm Cd$ and the microscopic interpretation of AMR in nuclei.

In general, rotational bands are built on intrinsic states violating rotational symmetry. In the common cases this symmetry is broken by a substantial quadrupole deformation
and therefore these bands show strong electric quadrupole ($E2$) transitions between the rotational states. Many exciting phenomena
have been discovered and predicted in this field such as backbending~\cite{Johnson1971Phys.Lett.605},
alignment phenomena~\cite{Stephens1972Nucl.Phys.257,Banerjee1973Nucl.Phys.366}, and superdeformed rotational bands~\cite{Twin1986Phys.Rev.Lett.811}. Such bands are most commonly interpreted as arising from the coherent collective rotation of many nucleons around an axis perpendicular to the symmetry axis
of the deformed density distribution~\cite{Bohr1975B}.

Over the past decades, however, a new type of rotational band with strong magnetic dipole ($M1$) and very weak $E2$ transitions has been discovered experimentally in nearly spherical light lead isotopes and other groups of nuclei (for reviews see Refs.~\cite{Hubel2005Prog.Part.Nucl.Phys.1,Frauendorf2001Rev.Mod.Phys.463,Clark2000Annu.Rev.Nucl.Part.Sci.1}). The intriguing feature here is that the orientation of the rotors is not specified by the deformation of the overall density but rather by the current distribution induced by specific nucleons moving in high-$j$ orbitals.

The explanation of such bands in terms of the ``shears mechanism'' was firstly given in Ref.~\cite{Frauendorf1993Nucl.Phys.A259}, where a decreasing tendency of
$B(M1)$ values with total angular momentum was also predicted. In this interpretation, the angular momentum vectors of the high-$j$ proton and neutron form two blades of a pair of shears and are almost perpendicular to each other at the bandhead. Along the bands, energy and angular momentum are increased by closing the blades of the shears, i.e., by aligning the proton and neutron angular momenta. Consequently, rotational bands are formed in spite of the fact that the shapes of these nuclei stay nearly spherical. A clear evidence for this new rotation mode has firstly been provided through lifetime measurements for four $M1$ bands in $^{198,199}\rm Pb$~\cite{Clark1997Phys.Rev.Lett.1868}. In order to distinguish this kind of rotation from the usual collective rotation in well-deformed nuclei (called electric rotation), the name ``magnetic rotation''~\cite{Frauendorf199452} was introduced. It alludes to the fact that here the magnetic moment is the order parameter inducing a violation of rotational symmetry and thus causing rotational-like structures in the spectrum~\cite{Frauendorf1997Z.Phys.A}. This forms an analogy to a ferromagnet where the total magnetic moment, the sum of the atomic dipole moments, is the order parameter.

In an antiferromagnet, on the other hand, one-half of the atomic dipole moments are aligned on one sublattice and
the other half are aligned in the opposite direction on the other sublattice. In such a way, the net magnetic moment in an antiferromagnet is canceled out. However, it is still an ordered state since the isotropy of such a state is also broken like a ferromagnet.

In analogy with an antiferromagnet, a similar phenomenon can be predicted in nuclei for ``antimagnetic rotation'' (AMR) in Ref.~\cite{Frauendorf1996272,Frauendorf2001Rev.Mod.Phys.463}: in specific nearly spherical nuclei, subsystems of valence protons (neutrons) are aligned back to back in opposite directions and nearly perpendicular to the orientation of the total spin of the valence neutrons (protons) (see Fig.~\ref{fig8}). Such an arrangement of the proton and neutron angular momenta also breaks rotational symmetry in these nearly spherical nuclei and causes excitations with rotational character on top of this bandhead. Along this band, energy and angular momentum are increased by simultaneous closing of the two blades of protons and neutrons toward the total angular momentum vector. Consequently, a new kind of rotational bands in nearly spherical nuclei is found showing some analogy with an antiferromagnet in solid state physics.

AMR is expected to be observed in the same regions in the nuclear chart as magnetic rotation~\cite{Frauendorf2001Rev.Mod.Phys.463}. However, it differs from magnetic rotation in two aspects. First, there are no $M1$ transitions in the AMR band since the transverse magnetic moments of the two subsystems are antialigned and canceled out against each other. Second, as the antimagnetic rotor is symmetric with respect to a rotation by 180$^\circ$ about the angular momentum axis, the energy levels in the AMR band differ in spin by $2\hbar$ and are connected by weak $E2$ transitions reflecting the nearly spherical core. Moreover, the phenomenon of AMR is characterized by a decrease of the $B(E2)$ values with spin, which has been demonstrated by lifetime measurements~\cite{Simons2003Phys.Rev.Lett.162501}. Since AMR was proposed~\cite{Frauendorf2001Rev.Mod.Phys.463}, it has attracted more and more attention both experimentally and theoretically. To date, experimental evidences of AMR have been reported in Cd isotopes including
$^{105}\rm Cd$~\cite{Choudhury2010Phys.Rev.C61308}, $^{106}\rm Cd$~\cite{Simons2003Phys.Rev.Lett.162501}, $^{108}\rm Cd$~\cite{Simons2005Phys.Rev.C24318,Datta2005Phys.Rev.C41305}, and $^{110}\rm Cd$~\cite{Roy2011Phys.Lett.B322}. In addition, the occurrence of this phenomenon in $^{109}\rm Cd$~\cite{Chiara2000Phys.Rev.C34318}, $^{100}\rm Pd$~\cite{Zhu2001Phys.Rev.C41302}, $^{144}\rm Dy$~\cite{Sugawara2009Phys.Rev.C64321} still needs further investigations.

Theoretically, the AMR bands have been discussed in simple geometry~\cite{Simons2003Phys.Rev.Lett.162501,Choudhury2010Phys.Rev.C61308} as well as the TAC model~\cite{Frauendorf2001Rev.Mod.Phys.463,Chiara2000Phys.Rev.C34318,Simons2003Phys.Rev.Lett.162501,Simons2005Phys.Rev.C24318,Zhu2001Phys.Rev.C41302}.
The TAC model can explicitly construct classical vector diagrams showing the angular momentum composition. This is of great help in visualizing the structure of the rotational bands. The quality of the cranking approximation for principal axis cranking~\cite{Meng1993ActaPhys.Sin.372}, tilted axis cranking~\cite{Frauendorf1996Z.Phys.A263}, and aplanar tilted axis cranking~\cite{Frauendorf1997Nucl.Phys.A131} has been discussed and tested within the particle rotor model. Based on the TAC model, many applications have been carried out in the framework of the pairing plus quadrupole model~\cite{Frauendorf2001Rev.Mod.Phys.463,Chiara2000Phys.Rev.C34318} or the microscopic-macroscopic model~\cite{Simons2003Phys.Rev.Lett.162501,Simons2005Phys.Rev.C24318,Zhu2001Phys.Rev.C41302}.

In these investigations, however, the polarization effects which are expected to have a strong influence on the quadrupole moments and thus the $B(E2)$ values are either neglected completely or taken into account only partially by minimizing the rotating energy surface with respect to a few deformation parameters. Moreover, the nuclear currents, which are the origin of symmetry violation in nuclei with AMR, are not treated in a self-consistent way in these models. Therefore, it is evident that a full understanding of AMR requires self-consistent and microscopic investigations including all degrees of freedom and based on reliable theories without additional parameters. Such calculations are not simple, but they are nowadays feasible in the framework of density functional theories.

During the past decades, the CDFT has received wide attention due to its success in
describing many nuclear phenomena in stable as well as in exotic nuclei~\cite{Ring1996Prog.Part.Nucl.Phys.193,Vretenar2005Phys.Rep.101,Meng2006Prog.Part.Nucl.Phys.470}. On the basis of universal functionals and without any additional parameters, the rotational excitations can be described in practical applications within the self-consistent cranking relativistic mean-field (RMF) framework. In rotating nuclei, time reversal symmetry is broken by the Coriolis operator and by unpaired nucleons. This leads to time-odd components in the fields.
In non-relativistic density functionals, these time-odd components are usually not very well defined, because the phenomenological parameters of these functionals are determined by experimental ground state properties in even-even nuclei where these time-odd fields vanish. In covariant density functionals this problem does not occur. The time-odd fields in the Dirac equation correspond to non-vanishing space-like components of the vector potential. They are often called {\it nuclear magnetism} and are induced by the various currents. Lorentz invariance requires that the time-like and the space-like components of the vector fields are determined by the same coupling constants, which are well determined by experimental ground state properties. Dobaczewski et al~\cite{Dobaczewski1995Phys.Rev.C1827} have emphasized that Galileian invariance induces constraints for the coupling constants in non-relativistic theories too. However, in many of the successful parameterizations of non-relativistic density functionals these conditions are not fulfilled and the spin parts of the functionals which are also connected with time-reversal breaking are not influenced by Galileian invariance. Therefore, it seems to us very important to use relativistic functionals for the description of magnetic and antimagnetic rotational bands, where the symmetry breaking mechanism is introduced by the currents.

A code for the cranked RMF model with arbitrary orientation of the rotational axis, i.e., three-dimensional cranking, has been developed in Ref.~\cite{Madokoro2000Phys.Rev.C61301}. However because of the numerical complexity of this code, so far, this model has been applied only for the magnetic rotation in $^{84}\rm Rb$. Focusing on more investigations of bands with magnetic rotation, a completely new computer code for self-consistent two-dimensional cranked RMF theory has been developed in Ref.~\cite{Peng2008Phys.Rev.C24313}. It includes significant improvements in implanting the simplex symmetry ${\cal P}_y{\cal T}$, quantum number transformation, and orientation constraints. Very recently, the TAC model based on a relativistic point-coupling Lagrangian which allows considerable simplification was established and applied successfully to the magnetic rotation both in light nuclei such as $^{60}\rm Ni$~\cite{Zhao2011Phys.Lett.B181} and in heavy nuclei such as $^{198,199}\rm Pb$~\cite{Yu2012Phys.Rev.C24318}.

The first fully self-consistent and microscopic investigation of AMR has been reported recently
in the framework of TAC based on a relativistic point-coupling Lagrangian in CDFT in Ref.~\cite{Zhao2011Phys.Rev.Lett.122501}. The present paper focuses on more detailed results in $^{105}\rm Cd$ including the Routhians, energies, angular momenta, moments of inertia, electromagnetic transitions, as well as the nuclear currents and densities. Moreover, the microscopic interpretation of AMR in nuclei such as the configuration, the characteristics of AMR, the ``two-shears-like'' mechanism, and the alignment of the angular momenta are discussed in detail. In Sec. II, the TAC formalism is implemented in CDFT using a point-coupling Lagrangian. The numerical details are given in Sec. III. In Sec. IV, a series of calculated results of the AMR band in $^{105}\rm Cd$ are presented. Finally, a summary is given in Sec. V.

\section{Theoretical framework}
The basic building blocks of point-coupling vertices are
\begin{equation}
    (\bar\psi{\cal O}\Gamma\psi),~~~~~{\cal O}\in\{1,\vec{\tau}\},~~~~~\Gamma\in\{1,\gamma_\mu,\gamma_5,
    \gamma_5\gamma_\mu,\sigma_{\mu\nu}\},
\end{equation}
where $\psi$ is the Dirac spinor field of nucleon, $\vec{\tau}$ is the vector of isospin Pauli matrices, and $\Gamma$
generally denotes $4\times4$ Dirac matrices. Throughout this paper, vectors in isospin space are denoted by arrows and space vectors by bold type. Greek indices $\mu$ and $\nu$ run over the Minkowski indices $0$, $1$, $2$, and $3$.

A general effective Lagrangian can be written as a power series in $\bar\psi{\cal O}\Gamma\psi$ and their derivatives, with higher-order terms representing in-medium many-body correlations. In the actual application we start with the following Lagrangian density of the form:
\begin{eqnarray}\label{Eq.Lag}
{\cal L}&=&{\cal L}^{\rm free}+{\cal L}^{\rm 4f}+{\cal L}^{\rm hot}+{\cal L}^{\rm der}+{\cal L}^{\rm em},\\
        &=&\bar\psi(i\gamma_\mu\partial^\mu-m)\psi \nonumber\\
        &&-\frac{1}{2}\alpha_S(\bar\psi\psi)(\bar\psi\psi)
          -\frac{1}{2}\alpha_V(\bar\psi\gamma_\mu\psi)(\bar\psi\gamma^\mu\psi)
        -\frac{1}{2}\alpha_{TV}(\bar\psi\vec{\tau}\gamma_\mu\psi)(\bar\psi\vec{\tau}\gamma^\mu\psi) \nonumber\\
        &&-\frac{1}{3}\beta_S(\bar\psi\psi)^3-\frac{1}{4}\gamma_S(\bar\psi\psi)^4-\frac{1}{4}
           \gamma_V[(\bar\psi\gamma_\mu\psi)(\bar\psi\gamma^\mu\psi)]^2 \nonumber\\
        &&-\frac{1}{2}\delta_S\partial_\nu(\bar\psi\psi)\partial^\nu(\bar\psi\psi)
          -\frac{1}{2}\delta_V\partial_\nu(\bar\psi\gamma_\mu\psi)\partial^\nu(\bar\psi\gamma^\mu\psi)
          -\frac{1}{2}\delta_{TV}\partial_\nu(\bar\psi\vec\tau\gamma_\mu\psi)\partial^\nu(\bar\psi\vec\tau\gamma_\mu\psi)\nonumber\\
        &&-\frac{1}{4}F^{\mu\nu}F_{\mu\nu}-e\frac{1-\tau_3}{2}\bar\psi\gamma^\mu\psi A_\mu,
\end{eqnarray}
which includes the Lagrangian density for free nucleons ${\cal L}^{\rm free}$, the four-fermion point-coupling terms ${\cal L}^{\rm 4f}$, the higher order terms ${\cal L}^{\rm hot}$ accounting for the medium effects, the derivative terms ${\cal L}^{\rm der}$ to simulate the effects of finite-range which are crucial for a quantitative description for nuclear density distributions (e.g., nuclear radii), and the electromagnetic interaction terms ${\cal L}^{\rm em}$. The higher order terms lead in the mean field approximation to density dependent coupling constants with a density dependence of polynomial form. Lagrangians with more general density dependencies~\cite{Niksic2008PhysRevC034318} can be used in a similar way.

In the tilted axis cranking approach based on CDFT, it is assumed that the nucleus rotates around an axis in the $xz$ plane and the Lagrangian in Eq.~(\ref{Eq.Lag}) is transformed into a frame rotating uniformly with a constant rotational frequency,
\begin{equation}
  \bm{\Omega}=(\Omega_x,0,\Omega_z)=(\Omega\cos\theta_\Omega,0,\Omega\sin\theta_\Omega),
\end{equation}
where $\theta_\Omega:=\sphericalangle(\bm{\Omega},\bm{e}_x)$ is the tilt angle between the cranking axis and the $x$ axis. From this rotating Lagrangian, the equation of motion for the nucleons can be derived in the same manner as in the meson-exchange case~\cite{Koepf1989Nucl.Phys.A61,Madokoro1997Phys.Rev.C2934} and one finds
 \begin{equation}\label{Eq.Dirac}
   [\bm{\alpha}\cdot(-i\bm{\nabla}-\bm{V})+\beta(m+S)
    +V^0-\bm{\Omega}\cdot\hat{\bm{J}}]\psi_k=\epsilon_k\psi_k,
 \end{equation}
where $\hat{\bm{J}}=\hat{\bm{L}}+\frac{1}{2}\hat{\bm{\Sigma}}$ is the total angular momentum of the nucleon spinors, and $\epsilon_k$ represents the single-particle Routhians for nucleons. The relativistic fields $S(\bm{r})$ and $V^\mu(\bm{r})$ read
\begin{subequations}\label{Eq.poten}
 \begin{eqnarray}
   S(\bm{r})&=&\alpha_S\rho_S+\beta_S\rho_S^2+\gamma_S\rho_S^3+\delta_S\triangle\rho_S, \\
   V^0(\bm{r})&=&\alpha_V\rho_V+\gamma_V\rho_V^3+\delta_V\triangle \rho_V+\tau_3\alpha_{TV} \rho_{TV}+\tau_3\delta_{TV}\triangle \rho_{TV}+eA^0, \\
   \bm{V}(\bm{r})&=&\alpha_V \bm{j}_V+\gamma_V(\bm{j}_V)^3+\delta_V\triangle \bm{j}_V+\tau_3\alpha_{TV} \bm{j}_{TV}+\tau_3\delta_{TV}\triangle \bm{j}_{TV}+e\bm{A},
 \end{eqnarray}
\end{subequations}
with $e$ the electric charge unit vanishing for neutrons. As usual, it is assumed that the nucleon single-particle
states do not mix isospin, i.e., the single-particle states are eigenstates of $\tau_3$. Therefore only the third component of isovector potentials survives. The Coulomb field $A^0(\bm{r})$ is determined by Poisson's equation
\begin{equation}
  -\triangle A^0(\bm{r}) = e\rho_c.
\label{Laplace}
\end{equation}
From prior experience in the meson-exchange case~\cite{Koepf1989Nucl.Phys.A61,Koepf1990Nucl.Phys.A279}, the spatial components of the electromagnetic vector potential $\bm{A}(\bm{r})$ are neglected since these contributions are extremely small.

In comparison with the coupled equations of motion in the case of meson-exchange potentials, one finds in the point-coupling model considerable simplifications. In particular, apart from the Laplace equation~(\ref{Laplace}), one does not have to solve the Klein-Gordon equations for mesons.

Since the Coriolis term $\bm{\Omega}\cdot\hat{\bm{J}}$ in the Dirac equation~(\ref{Eq.Dirac}) breaks time reversal symmetry in the intrinsic frame, currents are induced and as a consequence spatial components of the vector potential $\bm{V}(\bm{r})$. The densities and currents in Eqs.~(\ref{Eq.poten}) have the form
\begin{subequations}
 \begin{eqnarray}\label{currents}
   \rho_S(\bm{r})     &=& \sum_{i=1}^A \bar\psi_i(\bm{r})\psi_i(\bm{r})        , \\
   \rho_V(\bm{r})     &=& \sum_{i=1}^A \psi_i^\dagger(\bm{r})\psi_i(\bm{r})        , \\
   \bm{j}_V(\bm{r})   &=& \sum_{i=1}^A \psi_i^\dagger(\bm{r})\bm{\alpha}\psi_i(\bm{r}) , \\
   \rho_{TV}(\bm{r})  &=& \sum_{i=1}^A \psi_i^\dagger(\bm{r})\tau_3\psi_i(\bm{r})      , \\
   \bm{j}_{TV}(\bm{r})&=& \sum_{i=1}^A \psi_i^\dagger(\bm{r})\bm{\alpha}\tau_3\psi_i(\bm{r})    , \\
   \rho_{c}(\bm{r})   &=& \sum_{i=1}^A \psi_i^\dagger(\bm{r})\frac{1-\tau_3}{2}\psi_i(\bm{r}).
 \end{eqnarray}
\end{subequations}
Following, as usual, no-sea approximation the sums are taken over the particles states in the Fermi sea only, i.e., the contribution of the negative-energy states are neglected. In fact these contributions resulting in vacuum polarization are not neglected completely because this effect is taken to account in a global way by the adjustment of the parameters in the Lagrangian to experimental data~\cite{Zhu1991Phys.Lett.B325}.

By solving the equation of motion iteratively, one finally obtains the total energy in the laboratory frame
\begin{equation}
  \label{Eq.Etot}
   E_{\rm tot} = E_{\rm kin} + E_{\rm int} + E_{\rm cou} + E_{\rm c.m.},
\end{equation}
which is composed of a kinetic part
\begin{equation}
   E_{\rm kin} = \int d^3\bm{r}\sum\limits_{i=1}^A\psi_i^\dagger[\bm{\alpha}\cdot\bm{p}+\beta m]\psi_i,
\end{equation}
an interaction part
\begin{eqnarray}
    E_{\rm int} &=&\int d^3\bm{r} \left\{\frac{1}{2}\alpha_S\rho_S^2+\frac{1}{3}\beta_S\rho_S^3+\frac{1}{4}\gamma_S\rho_S^4
    +\frac{1}{2}\delta_S\rho_S\Delta\rho_S\right.\nonumber \\
    &&+\frac{1}{2}\alpha_V(\rho_V^2-\bm{j}\cdot\bm{j})
    +\frac{1}{2}\alpha_{TV}(\rho_{TV}^2-\bm{j}_{TV}\cdot\bm{j}_{TV})\nonumber \\
    &&+\frac{1}{4}\gamma_V(\rho_V^2-\bm{j}\cdot\bm{j})^2
    +\frac{1}{2}\delta_V(\rho_V\Delta\rho_V-\bm{j}\Delta\bm{j})\nonumber \\
    &&+\left.\frac{1}{2}\delta_{TV}(\rho_{TV}\Delta\rho_{TV}-\bm{j}_{TV}\Delta\bm{j}_{TV})\right\},
\end{eqnarray}
an electromagnetic part
\begin{equation}
    E_{\rm cou} = \int d^3\bm{r}\frac{1}{2}eA_0\rho_c,
\end{equation}
and the center-of-mass (c.m.) correction energy $E_{\rm c.m.}$ accounting for the treatment of center-of-mass motion
\begin{equation}
 E^{\rm mic}_{\rm c.m.}=-\frac{1}{2mA}\langle\hat{\bm P}^{2}_{\rm c.m.}\rangle,
\end{equation}
where $A$ is the mass number and $\hat{\bm{P}}_{\rm c.m.}=\sum_i^A\hat{\bm{p}}_i$ is the total momentum in the center-of-mass frame. It has been shown that this c.m. correction based on the self-consistently determined wave functions provides more reasonable and reliable results than that based on the simple oscillator model~\cite{Bender2000Euro.Phys.J.A467,Long2004Phys.Rev.C34319,Zhao2009Chin.Phys.Lett.112102}.

For each rotational frequency $\Omega$ the expectation values of the angular momentum components $\bm{J}=(J_x,J_y,J_z)$
in the intrinsic frame are given by
\begin{subequations}
\begin{eqnarray}
    J_x&=&\langle\hat{J}_x\rangle=\sum\limits_{i=1}^Aj^{(i)}_x, \\
    J_y&=& 0, \\
    J_z&=&\langle\hat{J}_z\rangle=\sum\limits_{i=1}^Aj^{(i)}_z,
\end{eqnarray}
\end{subequations}
and by means of the semiclassical cranking condition
\begin{equation}
 J = \sqrt{\langle\hat{J}_x\rangle^2+\langle\hat{J}_z\rangle^2}\equiv\sqrt{I(I+1)},
\end{equation}
one can the relate absolute value of the rotational frequency $\Omega$ to the angular momentum quantum number $I$ in the rotational band.

The orientation of the angular momentum vector $\bm{J}$ is represented by the angle $\theta_J:=\sphericalangle(\bm{J},\bm{e}_x)$ between the angular momentum vector $\bm{J}$ and the $x$ axis. In a fully self-consistent calculation, the orientation $\theta_J$ of the angular momentum $\bm{J}$ should be identical to the orientation  $\theta_\Omega$ of the angular velocity $\bm{\Omega}$.

The quadrupole moments $Q_{20}$ and $Q_{22}$ are calculated by
\begin{subequations}
\begin{eqnarray}
    Q_{20}&=&\sqrt{\frac{5}{16\pi}}\langle3z^2-r^2\rangle, \\
    Q_{22}&=&\sqrt{\frac{15}{32\pi}}\langle x^2-y^2\rangle,
\end{eqnarray}
\end{subequations}
and the deformation parameters $\beta$ and $\gamma$ can thus be extracted from
\begin{subequations}
\begin{eqnarray}
    \beta&=&\sqrt{a_{20}^2+2a_{22}^2}, \\
   \gamma&=&\arctan\left[\sqrt{2}\frac{a_{22}}{a_{20}}\right],
\end{eqnarray}
\end{subequations}
by using the relations
\begin{subequations}
\begin{eqnarray}
    Q_{20}&=&\frac{3A}{4\pi}R_0^2a_{20}, \\
    Q_{22}&=&\frac{3A}{4\pi}R_0^2a_{22},
\end{eqnarray}
\end{subequations}
with $R_0 = 1.2A^{1/3}~\rm fm$. Note that the sign convention in Ref.~\cite{Ring1980} is adopted for the definition of $\gamma$ here.

The nuclear magnetic moment is, in units of the nuclear magneton, given by~\cite{Yao2006Phys.Rev.C24307}
\begin{equation}
   \bm{\mu} = \sum\limits_{i=1}^A\int d^3r\left[\frac{mc^2}{\hbar c}q\psi^\dagger_i(\bm{r})\bm{r}\times\bm{\alpha}\psi_i(\bm{r})+\kappa\psi^\dagger_i(\bm{r})\beta\bm{\Sigma}\psi_i(\bm{r})\right],
\end{equation}
where the charge $q$ ($q_p=1$ for protons and $q_n=0$ for neutrons) is given in units of $e$, $m$ the nucleon mass, and $\kappa$ the free anomalous gyromagnetic ration of the nucleon ($\kappa_p=1.793$ and $\kappa_n=-1.913$).

From the quadrupole moments and the magnetic moment, the $B(M1)$ and $B(E2)$ transition probabilities can be derived in semiclassical approximation
\begin{subequations}
\begin{eqnarray}
   B(M1) &=& \frac{3}{8\pi}\mu_{\bot}^2 =\frac{3}{8\pi}(\mu_x\sin\theta_J-\mu_z\cos\theta_J)^2,\\
   B(E2) &=& \frac{3}{8}\left[ Q^p_{20}\cos^2\theta_J+\sqrt{\frac{2}{3}}Q^p_{22}(1+\sin^2\theta_J)\right]^2,
\end{eqnarray}
\end{subequations}
where $Q^p_{20}$ and $Q^p_{22}$ corresponds to the quadrupole moments of protons.

As in the tilted axis cranking relativistic mean-field (TAC-RMF) model with meson-exchange interaction~\cite{Peng2008Phys.Rev.C24313}, the directions of the principal axes of the density distribution are fixed along the $x$, $y$, and $z$ axes by a quadratic constraint on $\langle \hat{Q}_{2-1}\rangle=0$ in order to eliminate the numerical instabilities which results from the freedom of rotations around the $y$ axis. More details on this treatment can be seen in Ref.~\cite{Peng2008Phys.Rev.C24313}.

\section{Numerical details}

In the present work, as in Ref.~\cite{Peng2008Phys.Rev.C24313}, the invariance of space reflection $\cal P$ and the combination of the time reversal and reflection in the $y$ direction ${\cal P}_y{\cal T}$ are used in the code for the solution of the TAC-RMF model with a point-coupling interaction. The Dirac equation for the nucleons is solved in an isotropic three-dimensional Cartesian harmonic oscillator basis with $N_f=10$ major shells and the oscillator frequency given by $\hbar\omega_0=41A^{-1/3}$. By increasing $N_f$ from 10 to 12, the changes of total energies and total quadrupole moments are within 0.1\% and 4\% for the ground state of the nucleus $^{105}\rm Cd$, respectively. Therefore, a basis of $10$ major oscillator shells is adopted in the present calculations. Moreover, the point-coupling interaction PC-PK1~\cite{Zhao2010Phys.Rev.C54319} is used for the Lagrangian  without any additional parameters and pairing correlations are neglected.

In the recent experiment of $^{105}\rm Cd$~\cite{Choudhury2010Phys.Rev.C61308}, a negative-parity band was reported in forming the yrast line above $I=23/2\hbar$. It was assumed that this band corresponds to three aligned neutrons with the configuration $\nu[h_{11/2}(g_{7/2})^2]$ coupled to a pair of $g_{9/2}$ proton holes. As in Ref.~\cite{Choudhury2010Phys.Rev.C61308}, in the present calculations the configuration is fixed at the bandhead: The odd neutron is kept in the lowest level of the $h_{11/2}$ shell. The remaining nucleons are treated self-consistently by filling the orbitals according to their energy from the bottom of potential. This automatically leads  to the configuration shown in Fig.~\ref{fig1}.

In order to keep the configurations unchanged with increasing rotational frequency $\Omega$, we adopt the same prescription as in Ref.~\cite{Peng2008Phys.Rev.C24313}:  starting from the blocked level $|\psi_i(\Omega)\rangle$ at frequency $\Omega$, we block for frequency $\Omega+\delta\Omega$ the orbit $j$ which maximizes the overlap  $\langle\psi_j(\Omega+\delta\Omega)|\psi_i(\Omega)\rangle$, i.e.,
\begin{equation}\label{para.trans.}
  \langle\psi_j(\Omega+\delta\Omega)|\psi_i(\Omega)\rangle = 1 + {\cal O}(\delta\Omega).
\end{equation}
In the practical calculations, the rotational frequency $\Omega$ is increased successively in steps of $\delta(\hbar\Omega)=0.1~\rm MeV$ and in each case the prescription Eq.~(\ref{para.trans.}) is applied for the occupation of the various levels. Of course, as we see from Fig.~\ref{fig1}, this is in our case equivalent to apply, for each $\Omega$, this prescription only for the odd neutron in the $h_{11/2}$ shell and to determine the occupation of the other levels in a self-consistent way by filling the levels from the bottom of the well.

\section{Results and discussion}

In this section we present the results of the TAC-RMF calculations for the nucleus $^{105}\rm Cd$, in particular the Routhians, energies, angular momenta, moments of inertia, electromagnetic transitions, the currents and densities. Moreover, the microscopic interpretation of AMR in nuclei such as the configuration, the characteristics of AMR, the two-shears-like mechanism, and the alignment of the angular momenta are discussed in detail.

\subsection{Routhians}
\begin{figure}[!htbp]
\includegraphics[width=8cm]{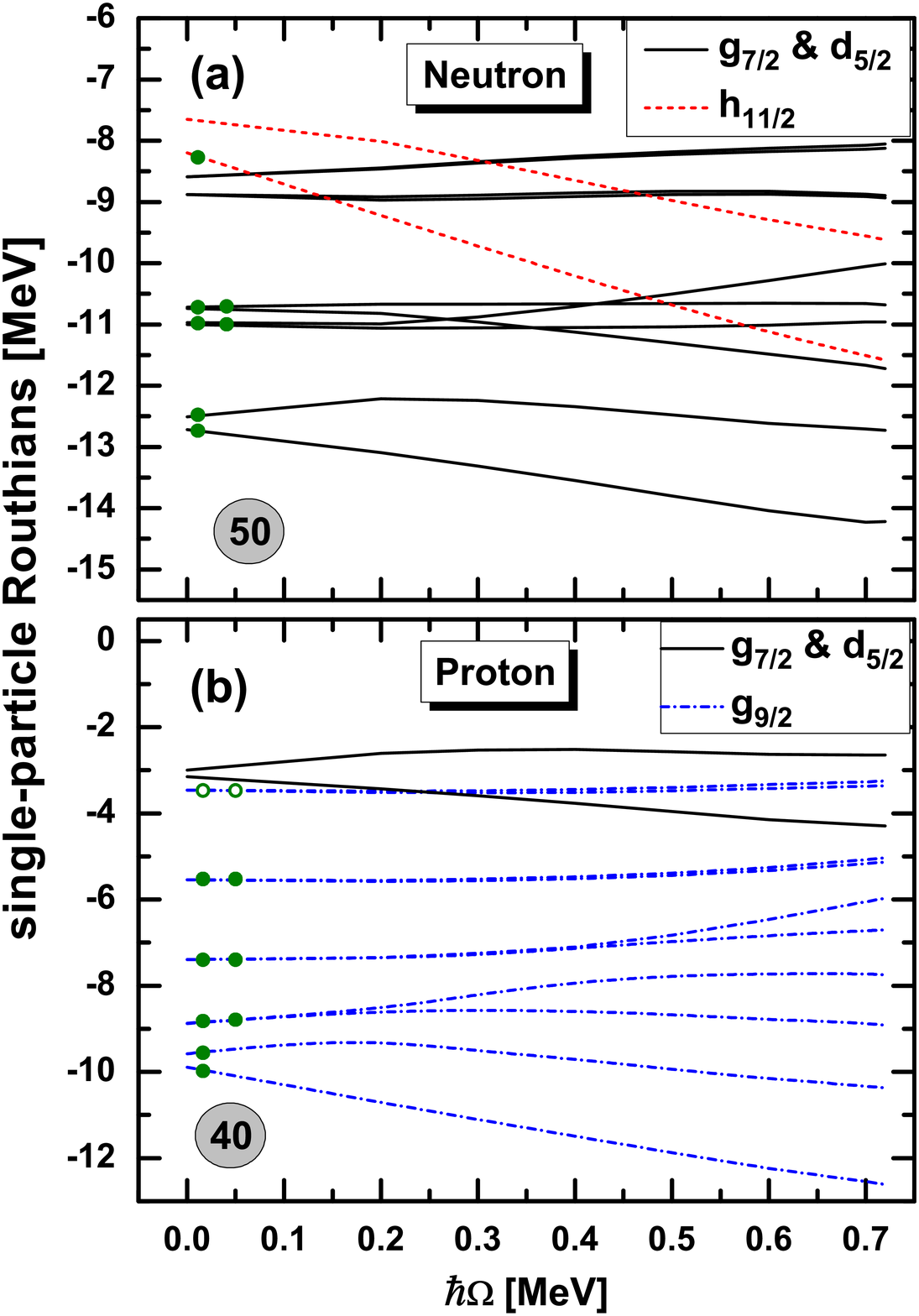}
\caption{(Color online) Single neutron (upper panel) and single proton (lower panel) Routhians near the Fermi surface in $^{105}\rm Cd$ as functions of the rotational frequency. The filled circles indicate the occupied levels, while the open circles in the lower panel represent the holes in the $g_{9/2}$ shell. }
\label{fig1}
\end{figure}

At finite cranking frequency $\Omega$, the Coriolis term violates time reversal invariance in the intrinsic frame and this leads to an energy splitting of the degenerate time reversal conjugate states at $\Omega=0$. This can be seen in Fig.~\ref{fig1} where the single particle Routhians $\epsilon_i$ in Eq.~(\ref{Eq.Dirac}) near the Fermi surface are shown as functions of the rotational frequency. The amplitudes of this energy splitting in $^{105}\rm Cd$ range from 0 to 2.3~MeV. As expected, we find a larger splitting for states with larger expectation values $|j_x|$. This results from the Coriolis term which reads $\Omega_xJ_x$ when the rotational axis is along the $x$ axis. It should be noted that even without Coriolis term, i.e. for zero angular velocity, time reversal symmetry is broken by the current of the odd particle. Therefore the energy levels are not exactly twofold degenerate and some of them show some splitting in the energy already at $\Omega=0$. This phenomenon has been discussed, e.g., in Ref.~\cite{Yao2006Phys.Rev.C24307}.

In Fig.~\ref{fig1}, the occupied single-particle levels are marked by filled circles. For the protons, one can distinctly see that two proton holes are always sitting at the top of the $g_{9/2}$ shell. For the neutrons, there are totally seven valence neutrons above the $N=50$ shell. As discussed in the previous section, the odd neutron is kept fixed in the lowest level in the $h_{11/2}$ shell. The other six remaining neutrons are treated self-consistently by filling the orbitals according to their energy and distributed over the $(g_{7/2}d_{5/2})$ shell. We find that there is a strong mixing between these orbitals. In this sense, we have here, in the present microscopic calculations, a similar configuration as that of Ref.~\cite{Choudhury2010Phys.Rev.C61308}.

\begin{figure}[!htbp]
\includegraphics[width=8cm]{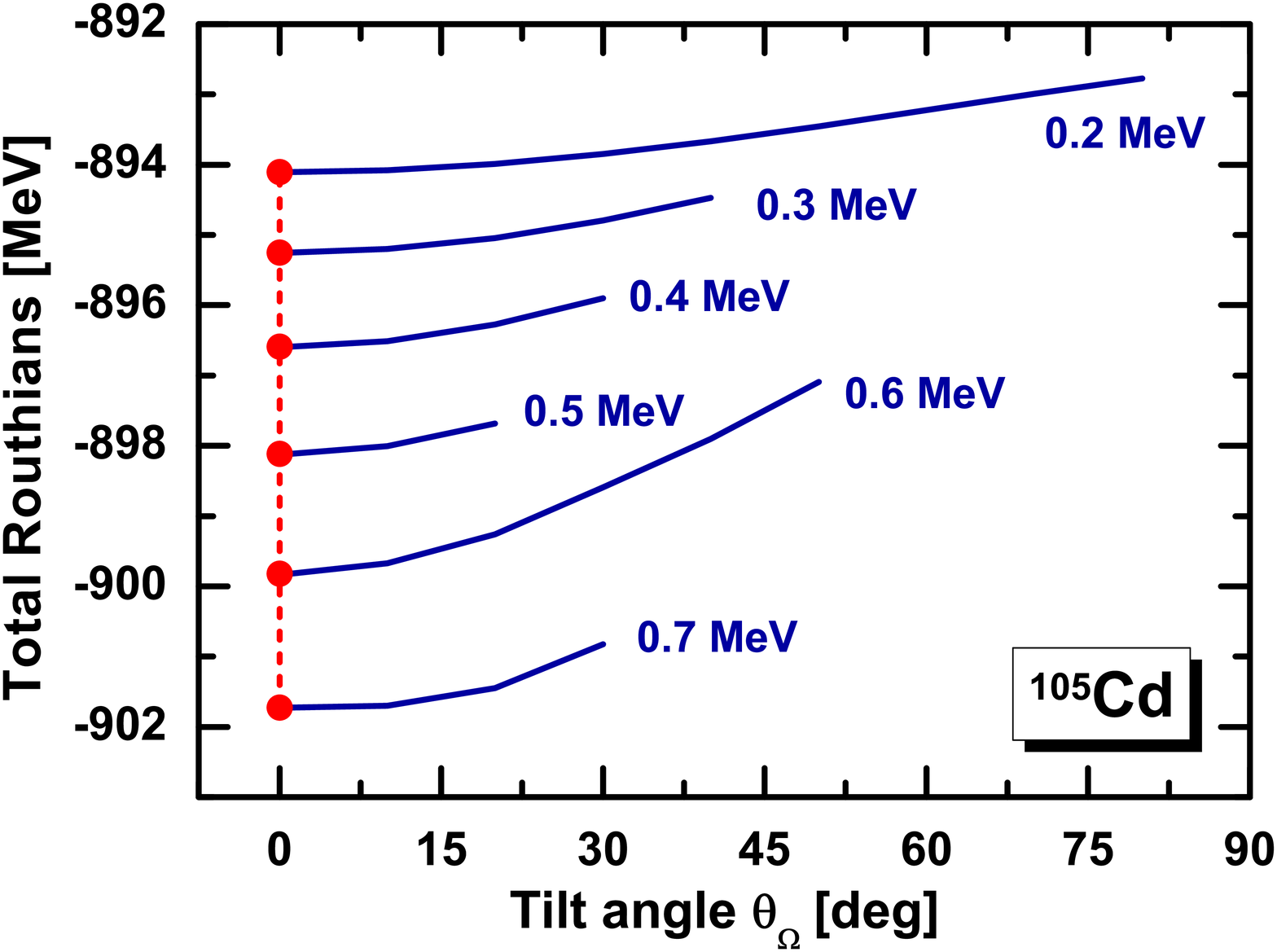}
\caption{(Color online) TAC-RMF results for total Routhians (solid lines) as functions of the tilt angle $\theta_\Omega$. The rotational frequencies range from $\hbar\Omega = 0.2$~MeV to 0.7~MeV. The solid dots represent the minima for the corresponding frequency.}
\label{fig2}
\end{figure}

So far we kept the tilt angle $\theta_\Omega$ always fixed at $\theta_\Omega=0$. Of course, this angle should be determined in a self-consistent way by minimizing the total Routhian
\begin{equation}
  E'(\Omega,\theta_\Omega) = E_{\rm tot}-\langle \cos\theta_\Omega\Omega J_x+\sin\theta_\Omega\Omega J_z \rangle
\end{equation}
with respect to the angle $\theta_\Omega$. Therefore we repeated this calculations for a range of $\theta_\Omega$ values and in Fig.~\ref{fig2} we show the total Routhians as functions of the tilt angle $\theta_\Omega$ for rotational frequencies ranging from $\hbar\Omega = 0.2$~MeV to 0.7~MeV. It turns out that the minimum of the total Routhians (solid dot) remains for all angular velocities $\Omega$ at the tilt angle $\theta_\Omega = 0^\circ$. This demonstrates clearly that the rotational axis is always parallel to the $x$ axis in the present antimagnetic rotor. Of course this is a natural consequence of the symmetry for the antimagnetic configuration, but it is not trivial, because symmetries can be broken in the self-consistent solution. Of course, having this result we can conclude that, in principle, both the principal axis cranking (PAC) and TAC calculations can be used to describe AMR. The reasons for carrying out TAC calculations here is that the relative orientation of the $g_{9/2}$ proton hole vectors can be calculated easily (see Fig.~\ref{fig7} below).

\subsection{Energy and moment of inertia}

\begin{figure}[!htbp]
\includegraphics[width=8cm]{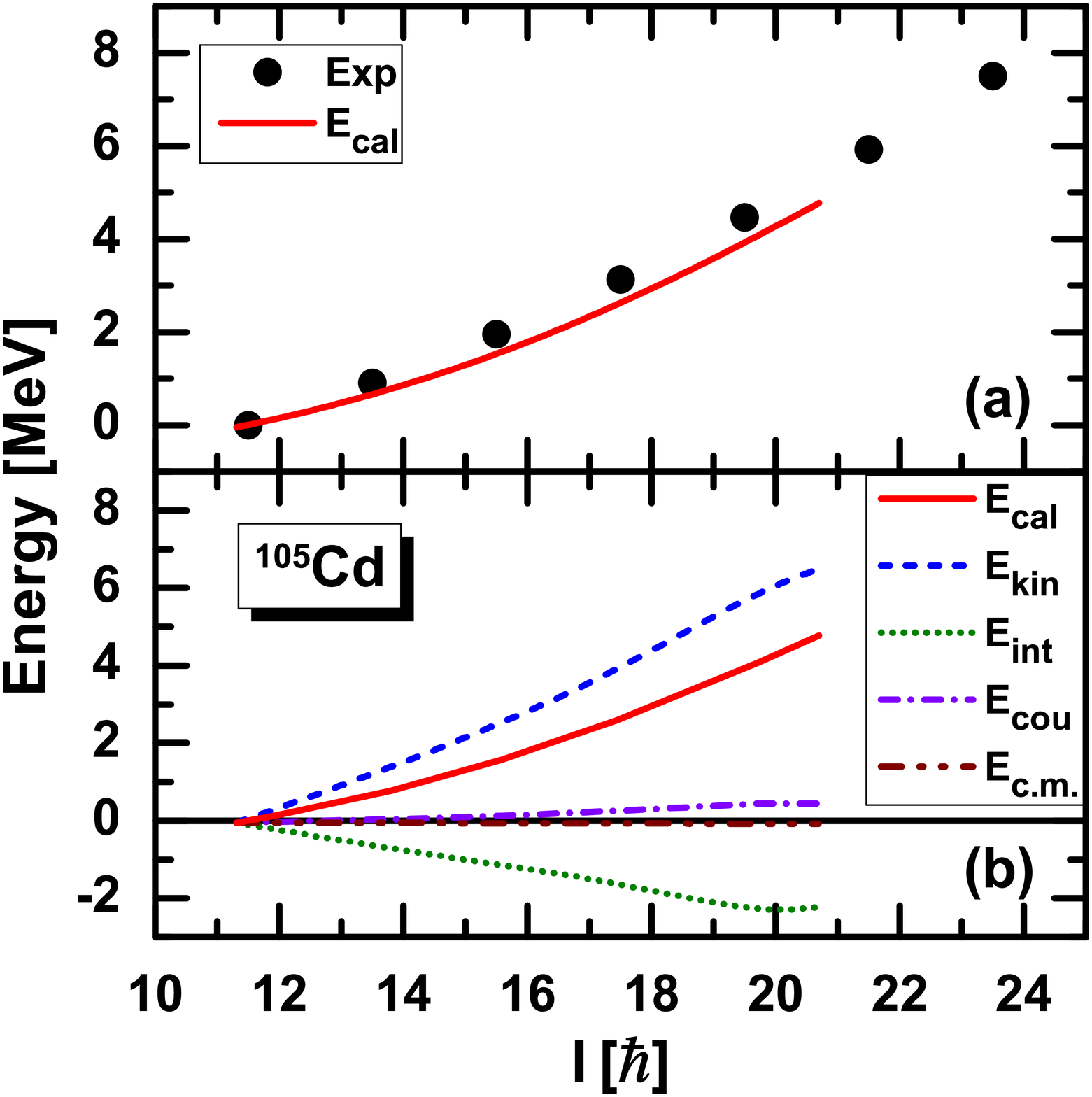}
\caption{(Color online) Energy spectrum obtained from the TAC-RMF calculations in comparison with the data (upper panel) as well as the corresponding contributions from kinetic energy, interaction energy, Coulomb energy, and center-of-mass correction energy (lower panel). These results are normalized at $I = 23/2\hbar$.}
\label{fig3}
\end{figure}

In the upper panel of Fig.~\ref{fig3}, the calculated total energy (full line) is shown in comparison with the data~\cite{Choudhury2010Phys.Rev.C61308}. The experimental energy spectrum is reproduced in an excellent way by the present self-consistent calculations. In Eq.~(\ref{Eq.Etot}), the calculated energy is divided into kinetic energy, interaction energy, Coulomb energy, and center-of-mass correction energy. The contributions to the total energy from these four parts are shown in the lower panel of Fig.~\ref{fig3}. It is found that the Coulomb energy and center-of-mass correction energy are barely changed in amplitude (less than 0.5~MeV).
In comparison, the kinetic energy increases monotonously up to roughly 6~MeV and the interaction energy declines monotonously up to roughly 2~MeV. Therefore we can conclude that the rotational excitation energy in the band comes mainly from the kinetic energy with a moderate competition from the interaction energy.

\begin{figure}[!htbp]
\includegraphics[width=8cm]{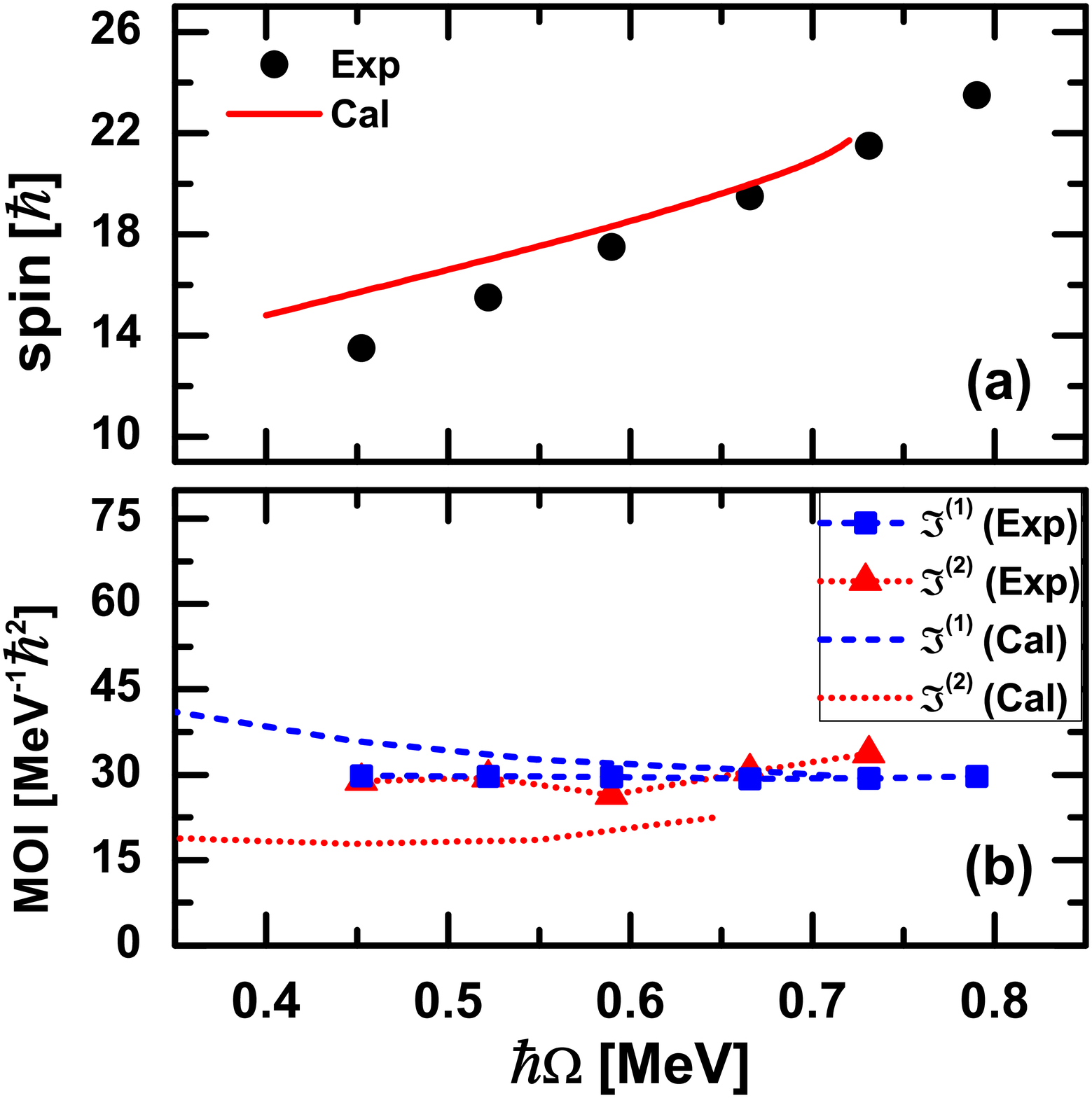}
\caption{(Color online) The total angular momentum (upper panel) as well as the kinetic ${\mathfrak{J}}^{(1)}$ and dynamic ${\mathfrak{J}}^{(2)}$ moments of inertia (lower panel) as functions of the rotational frequency in the TAC-RMF calculations in comparison with the data.}
\label{fig4}
\end{figure}

In Fig.~\ref{fig4}, the total angular momentum (upper panel) as well as the kinetic ${\mathfrak{J}}^{(1)}$ and dynamic ${\mathfrak{J}}^{(2)}$ moments of inertia (lower panel) are shown as functions of the rotational frequency and compared with the data. Taking into account the quantal corrections~\cite{Frauendorf1996Z.Phys.A263}, the total angular momentum $J$ calculated in TAC model corresponds to the quantum number of the angular momentum $I+1/2$ since $\sqrt{I(I+1)}\approx I+1/2$. This prescription permits us to compare the TAC calculations with the experimental data. In the present AMR case, states differing by two units of angular momentum are arranged into a $\Delta I =2$ band. Therefore, the experimental rotational frequency is extracted as
\begin{equation}
  J=I-1/2, \quad\quad\quad\hbar\Omega_{\rm exp} = \frac{1}{2} E_\gamma(I\rightarrow I-2).
\end{equation}
Meanwhile, the experimental values of kinetic ${\mathfrak{J}}^{(1)}$ and dynamic ${\mathfrak{J}}^{(2)}$ moments of inertia are calculated from the transition energies by using the finite difference approximation for $\Delta I =2$ bands,
\begin{subequations}
\begin{equation}
   {\mathfrak{J}}^{(1)} = \frac{2I}{E_\gamma(I\rightarrow I-2)},
\end{equation}
\begin{equation}
   {\mathfrak{J}}^{(2)} = \frac{4}{E_\gamma(I+2\rightarrow I)-E_\gamma(I\rightarrow I-2)},
\end{equation}
\end{subequations}
with $\hbar\Omega_{\rm exp} = \frac{1}{2} E_\gamma(I\rightarrow I-2)$.

In the upper panel of Fig.~\ref{fig4}, it can be found that the calculated total angular momenta increase almost linearly with increasing rotational frequency and agree with the data very well. This leads to fairly constant values for both the kinetic and dynamic moments of inertia as shown in the lower panel.
Generally, the fact that both the kinetic and the dynamic moments of inertia stay roughly constant is well reproduced. However, the calculated values for ${\mathfrak{J}}^{(2)}$ underestimate the data.
%The good agreement is obviously connected with the fact that the time-even and time-odd mean fields are coupled with the same coupling constants in CDFT due to Lorentz invariance, %whereas in many non-relativistic calculations the coupling constants of the time-odd parts are not adjusted properly.

\subsection{Transition probability $B(E2)$ and deformation}

\begin{figure}[!htbp]
\includegraphics[width=8cm]{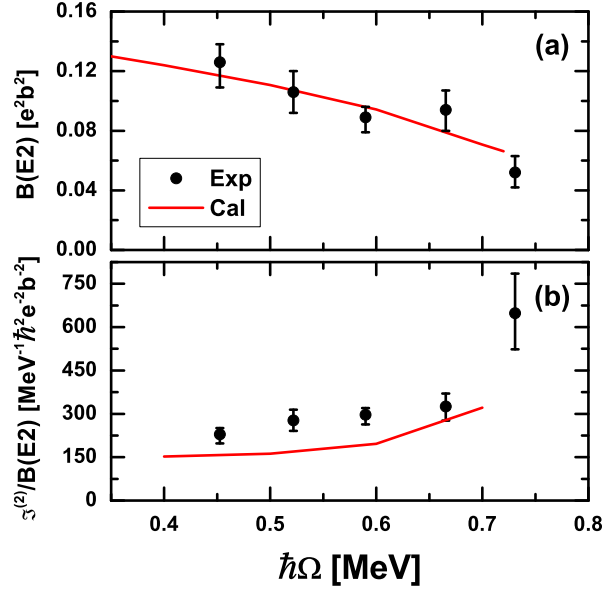}
\caption{(Color online) $B(E2)$ values (upper panel) and ${\mathfrak{J}}^{(2)}/B(E2)$ ratios (lower panel) as functions of the rotational frequency in the TAC-RMF calculation in comparison with the data. }
\label{fig5}
\end{figure}

Typical characteristics of AMR include weak $E2$ transitions, reflecting the small deformation of the core, which results in large ratios of the dynamic moments of inertia ${\mathfrak{J}}^{(2)}$ to the reduced transition probability $B(E2)$ values. What's more, the $B(E2)$ values rapidly decrease with the angular momentum.
In the upper and lower panels of Fig.~\ref{fig5}, the calculated $B(E2)$ values and the ${\mathfrak{J}}^{(2)}/B(E2)$ ratios are shown as functions of the rotational frequency in comparison with the data. It is found that the $B(E2)$ values are in very good and the ${\mathfrak{J}}^{(2)}/B(E2)$ ratios in reasonable agreement with the data. Quantitatively, one can see that the calculated $B(E2)$ values are small $(< 0.14~e^2b^2)$ and this leads to the large ${\mathfrak{J}}^{(2)}/B(E2)$ ratios $(> 150~{\rm MeV}^{-1}\hbar^2e^{-2}b^{-2})$. Moreover, the fact that the $B(E2)$ values decrease smoothly with the growing rotational frequency, which results in an increase of the ${\mathfrak{J}}^{(2)}/B(E2)$ ratios, is connected with the interpretation of the two-shears-like mechanism.

It should be noted that in some nuclei (e.g., $^{108}\rm Sn$ and $^{109}\rm Sb$) in the $A\sim110$ mass region~\cite{Afanasjev1999Phys.Rep.1}, a so-called ``smoothly terminating band'' is observed. Its structure gradually develops from a prolate shape rotating around an axis perpendicular to the symmetry axis, through a sequence of triaxial shapes, to an oblate shape with the symmetry axis parallel to the total angular momentum vector. In this sense, the band terminates when the angular momenta of the valence particles are fully aligned, just as the band of the antimagnetic rotor terminates when the shears blades are aligned. However, one of the most important feature for AMR is the rise of ${\mathfrak{J}}^{(2)}/B(E2)$ ratios reflecting the fact that ${\mathfrak{J}}^{(2)}$, as shown in the lower panel of Fig.~\ref{fig4}, is essentially constant whereas the $B(E2)$ values decline along the band. This is not the case for a smoothly terminating band which has an almost constant ${\mathfrak{J}}^{(2)}/B(E2)$ ratio, indicating that the ${\mathfrak{J}}^{(2)}$ have falling behavior as similar as $B(E2)$ with the increasing spin.

\begin{figure}[!htbp]
\includegraphics[width=7cm]{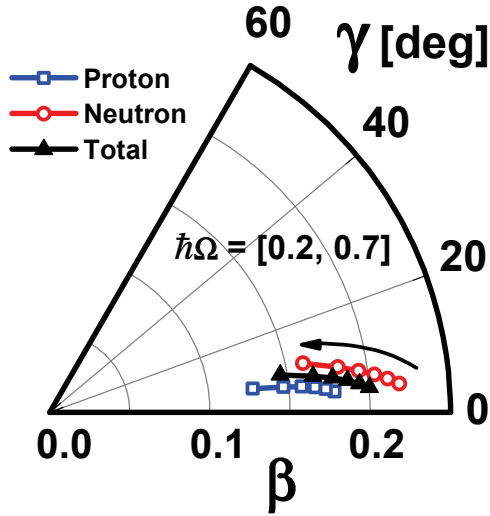}
\caption{(Color online) Evolutions of the neutron, proton and total deformations in the $(\beta, \gamma)$ plane with rotational frequency in the TAC-RMF calculations. The arrow indicates the increasing direction of rotational frequency.}
\label{fig6}
\end{figure}

The decrease of the $B(E2)$ values can be understood by the changes in the nuclear deformation.
In Fig.~\ref{fig6} we present the evolution of the deformation for neutrons, protons and the whole nucleus in the $(\beta, \gamma)$ plane with increasing rotational frequency.
It is found that the deformations for neutrons, protons and the whole nucleus behave in a similar way, i.e., a rapid decrease in $\beta$ together with a small and nearly constant triaxiality. In particular, the calculated proton deformation reflecting the charge distribution and therefore responsible for the $B(E2)$ values decreases from $\beta=0.18$ to $\beta=0.12$ with rather small triaxiality ($\gamma\le7^\circ$). This is related to the decrease of the $B(E2)$ values shown in the upper panel of Fig.~\ref{fig5} and one can thus conclude that the alignment of the shears blades in AMR, i.e., the two-shears-like mechanism, is accompanied by a transition from a prolate towards a nearly spherical shape.

\subsection{Two-shears-like mechanism}

\begin{figure}[!htbp]
\includegraphics[width=8cm]{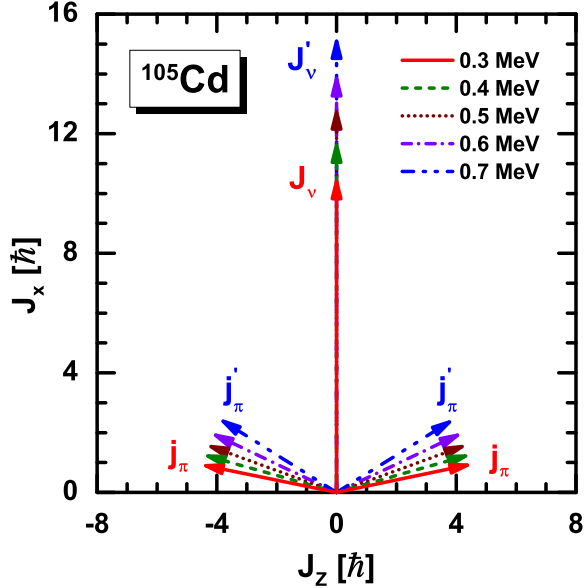}
\caption{(Color online) Angular momentum vectors of neutrons $J_\nu$ and the two $g_{9/2}$ proton holes $j_\pi$ at rotational frequencies from 0.3 to 0.7~MeV. }
\label{fig7}
\end{figure}

In order to examine the two-shears-like mechanism for the AMR band in $^{105}\rm Cd$, we show in Fig.~\ref{fig7} the angular momentum vectors of the neutrons $\bm{J}_\nu$ and the two $g_{9/2}$ proton-holes $\bm{j}_\pi$ for rotational frequencies ranging from 0.3 to 0.7~MeV. The neutron angular momentum $\bm{J}_\nu$ is defined as
 \begin{equation}
   \bm{J}_\nu=\langle\hat{\bm{J}}_\nu\rangle=\sum\limits_{n=1}^N\bm{j}^{(n)}_\nu=\sum\limits_{n=1}^N\langle n|\hat{\bm{J}}|n\rangle,
 \end{equation}
where the sum runs over all the neutron levels occupied in the cranking wave function in the intrinsic system. At the bandhead ($\hbar\Omega = 0.3~\rm MeV$), each of the two proton angular momentum vectors $\bm{j}_\pi$ are pointing opposite to each other and are nearly perpendicular to the vector $\bm{J}_\nu$. They form the blades of the two shears. As the rotational frequency increases, the gradual alignment of the vectors $\bm{j}_\pi$ of the two $g_{9/2}$ proton-holes toward the vector $\bm{J}_\nu$ generates angular momentum, while the direction of the total angular momentum stays unchanged. This leads to the closing of the two shears simultaneously by moving one blade towards the other. In such a way, the two-shears-like mechanism in $^{105}\rm Cd$ is clearly demonstrated in Fig.~\ref{fig7}. Similar results on the distribution of angular momenta for the $g_{9/2}$ proton holes and the $h_{11/2}$ neutron(s) in other Cd isotopes can also be obtained with TAC in the framework of either the pairing plus quadrupole model~\cite{Chiara2000Phys.Rev.C34318} or the microscopic-macroscopic model~\cite{Simons2003Phys.Rev.Lett.162501,Simons2005Phys.Rev.C24318}.

\begin{figure}[!htbp]
\includegraphics[width=10cm]{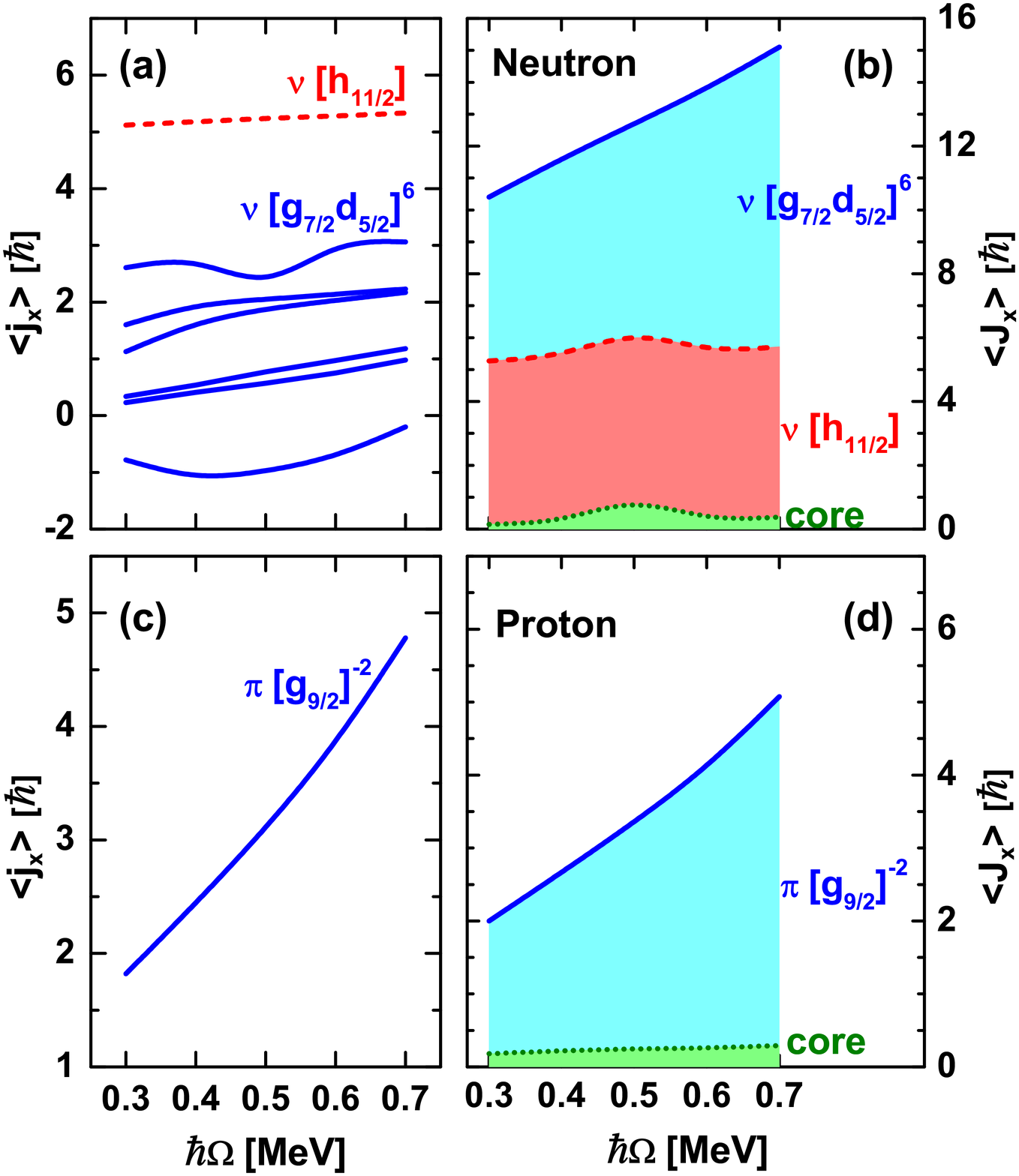}
\caption{(Color online) Angular momentum alignment for the seven neutron particles [panel (a)]  and two proton holes [panel (c)] as well as the compositions of the neutron [panel (b)] and proton [panel (d)] angular momenta.  }
\label{fig8}
\end{figure}

In a fully self-consistent and microscopic calculation, both the energy and angular momentum come from individual nucleons. In order to investigate this two-shear-like mechanism in a more microscopic way, we show in Fig.~\ref{fig8} the angular momentum alignment for the seven neutron particles [panel (a)] and two proton holes [panel (c)] as well as the compositions of the neutron [panel (b)] and proton [panel (d)] angular momenta.

It is found that the spin contributions from the proton and neutron cores, i.e., from the orbitals below $Z=50$ and $N=50$, are quite small. This $^{100}$Sn core stays relatively inert. Instead, the contributions to the angular momenta mainly arise from the high-$j$ orbitals, i.e., from proton holes in the $g_{9/2}$ shell as well as from neutron particles in the $h_{11/2}$ and $g_{7/2}$ shells.

For the protons, the angular momentum is mainly built from the two holes at the top of the $g_{9/2}$ shell while the spin contribution from the $Z=50$ core is rather small ($<0.5\hbar$). As shown in Fig.~\ref{fig7}, the two proton holes cancel each other in the $z$ direction. With growing angular velocity, the two proton holes align gradually along the rotational axis and the angular momentum in $x$ direction increases. The blades of the two shears are closing.

For the neutrons, the angular momentum along the $x$ axis originates mainly from the neutron particles filling in the bottom of the $h_{11/2}$ and $g_{7/2}$ shell whereas the contribution from the $N=50$ core is less than $1 \hbar$. A neutron sitting in the $h_{11/2}$ orbit contributes an angular momentum of roughly $5\hbar$ along the $x$ axis. As the rotational frequency increases, the contribution of this neutron does not change much and the increase of angular momentum is mostly generated by the alignment of the other six neutrons in the $g_{7/2}$ and $d_{5/2}$ orbitals.

The small bump shown in the spin contributions of the neutron core at $\hbar\Omega \sim 0.5~\rm MeV$ is associated with a virtual crossing between the core $g_{9/2}$ and the valence $(g_{7/2}d_{5/2})$ orbitals where the spin contributions can not be split properly between these crossing orbitals. As the level crossing involves only the filled orbitals, there is no influence on the total angular momentum but we see it only in the somewhat artificial division of spin contributions between the core and valence particles.

\begin{figure}[!htbp]
\includegraphics[width=11cm]{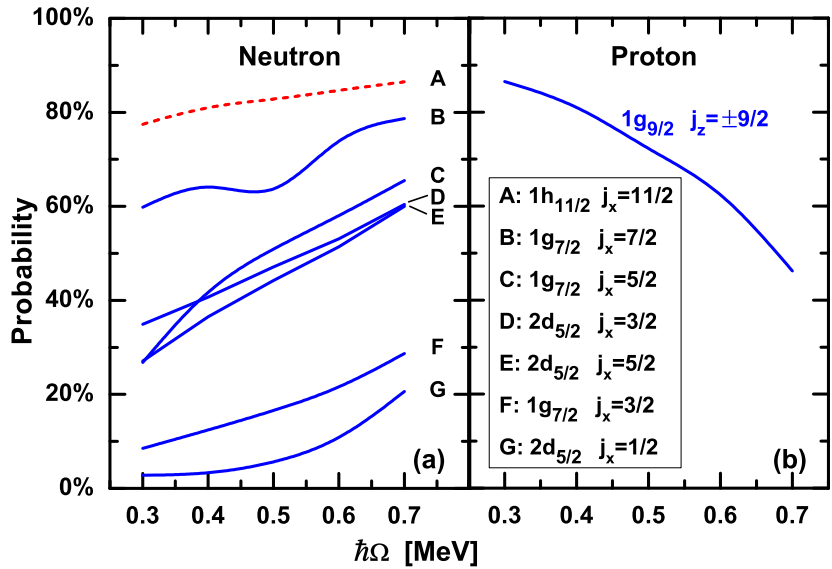}
\caption{(Color online) Probability variations as functions of the rotational frequency for the main components of the valence single-particle states at $\hbar\Omega=0.7$ MeV. The probabilities of $j_x$ for seven valence neutron orbitals and the probabilities of $j_z$ for two valence proton orbitals are displayed in the left and right panels respectively.}
\label{fig9}
\end{figure}

In the present microscopic calculation, when a configuration is labeled as having $n$ particles, e.g., in the ($g_{7/2}d_{5/2}$) subshells, it does not mean that the particles  in the deformed rotating potential are in these pure sub-shells, but instead they are distributed over spherical orbits with the quantum numbers ($ n\ell j j_x$), where $n$ is the radial quantum number, ($\ell j$) are the angular momentum quantum numbers and $j_x$ is the magnetic quantum number along the $x$-axis. Therefore, both the alignment of proton holes and neutron particles arise from a considerable mixing between the single-particle orbitals.

In order to quantify these statements somewhat we expand the wavefunction of the rotating particle $i$ in terms of the spherical oscillator basis states
\begin{equation}
\vert i\rangle = \sum_{n\ell j j_x} \langle n\ell j j_x| i\rangle \vert n\ell j j_x\rangle
\label{expansion}
\end{equation}
and show in Fig.~\ref{fig9} the probabilities $|\langle n\ell jj_x| i\rangle|^2$ that the orbit ($ n\ell j j_x$) (labeled by the letters A, B, C, D, E, F, G) is occupied by a particle $i$ in the following way: In the left panel we concentrate on the seven neutrons orbits in the valence shell. Starting with the maximally aligned situation at the angular velocity $\hbar\Omega=0.7$ MeV we calculate for each orbit ($ n\ell j j_x$) the particle $i$ with the highest probability  to sit in this shell. It turns out that the aligned $1h_{11/2},j_x=11/2$ orbit is occupied by the odd neutron with a very high probability of roughly 85 \%. In the rest of the valence shell there is more mixing. Of course, these probabilities depend on the angular velocity. Next we go to lower frequencies and determine for each orbit ($ n\ell j j_x$) the particle $i$ with the maximal overlap to the corresponding particle at the neighboring next higher frequency applying the rule given in Eq. (\ref{para.trans.}) and we find that there is not much change in the occupation of the aligned $1h_{11/2},j_x=11/2$ orbit. At the bandhead it is still occupied with a probability of 77 \%. As shown in Fig.~\ref{fig8} it always contributes $\sim5\hbar$ to the angular momentum along the $x$ axis. However, the other probabilities decrease with decreasing angular velocity. When we reach the bandhead, the corresponding probabilities have decreased by more than 20 percentage points. Starting from the bandhead, the increase of angular momentum is generated mostly by the alignment of the six neutrons in the $g_{7/2}$ and $d_{5/2}$ orbitals.

Next we consider the protons in the right panel of Fig.~\ref{fig9}. Here we show only the two $1g_{9/2}$ orbits with the angular momentum quantum numbers $j_z=\pm9/2$ along the $z$-axis. both probabilities are equal because of the symmetry with respect to a rotation of 180$^\circ$ around the $x$-axis. At the band head, at $\hbar\Omega=0.3$ MeV, the two protons holes occupy these orbitals with 86 \% probability, but with increasing angular velocity this occupation rapidly decreases because the holes align along the rotational $x$-axis. This alignment is characterized by a considerable mixing of orbitals with other $j_z$ values.

In short, the present microscopic calculation demonstrates that apart from the two proton holes in the $g_{9/2}$ orbital and one neutron particle in the $h_{11/2}$ orbital, the other neutrons distributed over several subshells above the $N=50$ core also contribute jointly to the AMR band in $^{105}\rm Cd$. The total angular momentum results from the alignment of the proton holes and the mixing within the neutron orbitals. Because of this strong mixing between the neutron orbitals, the phenomenological interpretation in Ref.~\cite{Choudhury2010Phys.Rev.C61308} without proper treatment of the core is not fully justified.

\subsection{Nucleon currents and densities}

\begin{figure}[!htbp]
\includegraphics[width=8cm]{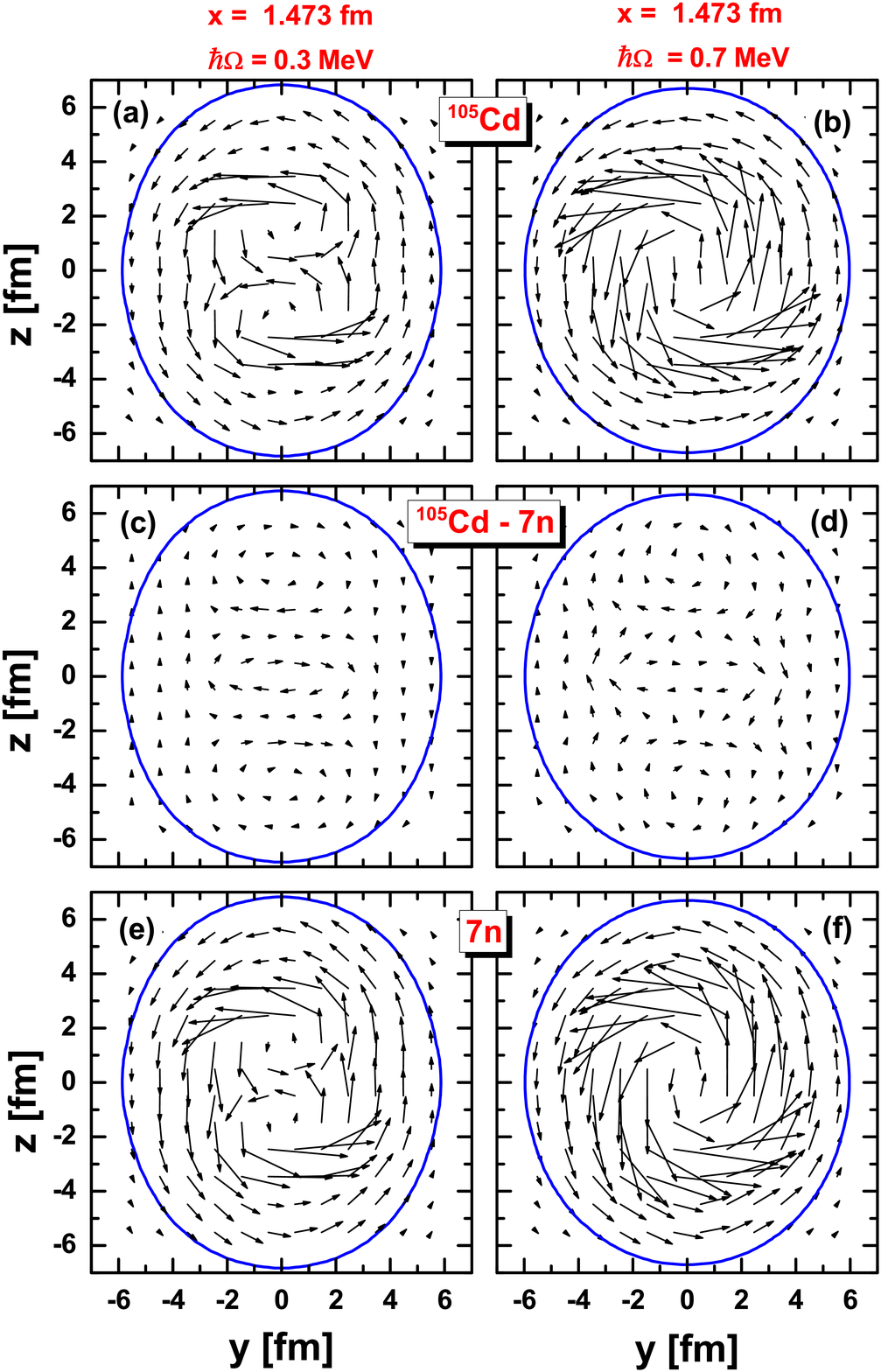}
\caption{(Color online)
Dirac currents for $^{105}\rm Cd$ in the $yz$ plane perpendicular to the rotational axis at $x=1.473~\rm fm$ for all neutrons (top), for the $N=50$ core (middle), and for the seven valence neutrons (bottom) at $\hbar\Omega=0.3~\rm MeV$ (left panels) and $\hbar\Omega=0.7~\rm MeV$ (right panels). The shape and the size of the nucleus $^{105}\rm Cd$ are represented by the solid lines corresponding to lines with constant neutron density $\rho_n=0.01~{\rm fm}^{-3}$.}
\label{fig10}
\end{figure}

As discussed in the introduction, the time reversal symmetry is broken in rotating nuclei. This leads to polarization currents and nuclear magnetism, i.e., time-odd components of the fields. In the following, we will investigate the current distributions.

\begin{figure}[!htbp]
\includegraphics[width=8cm]{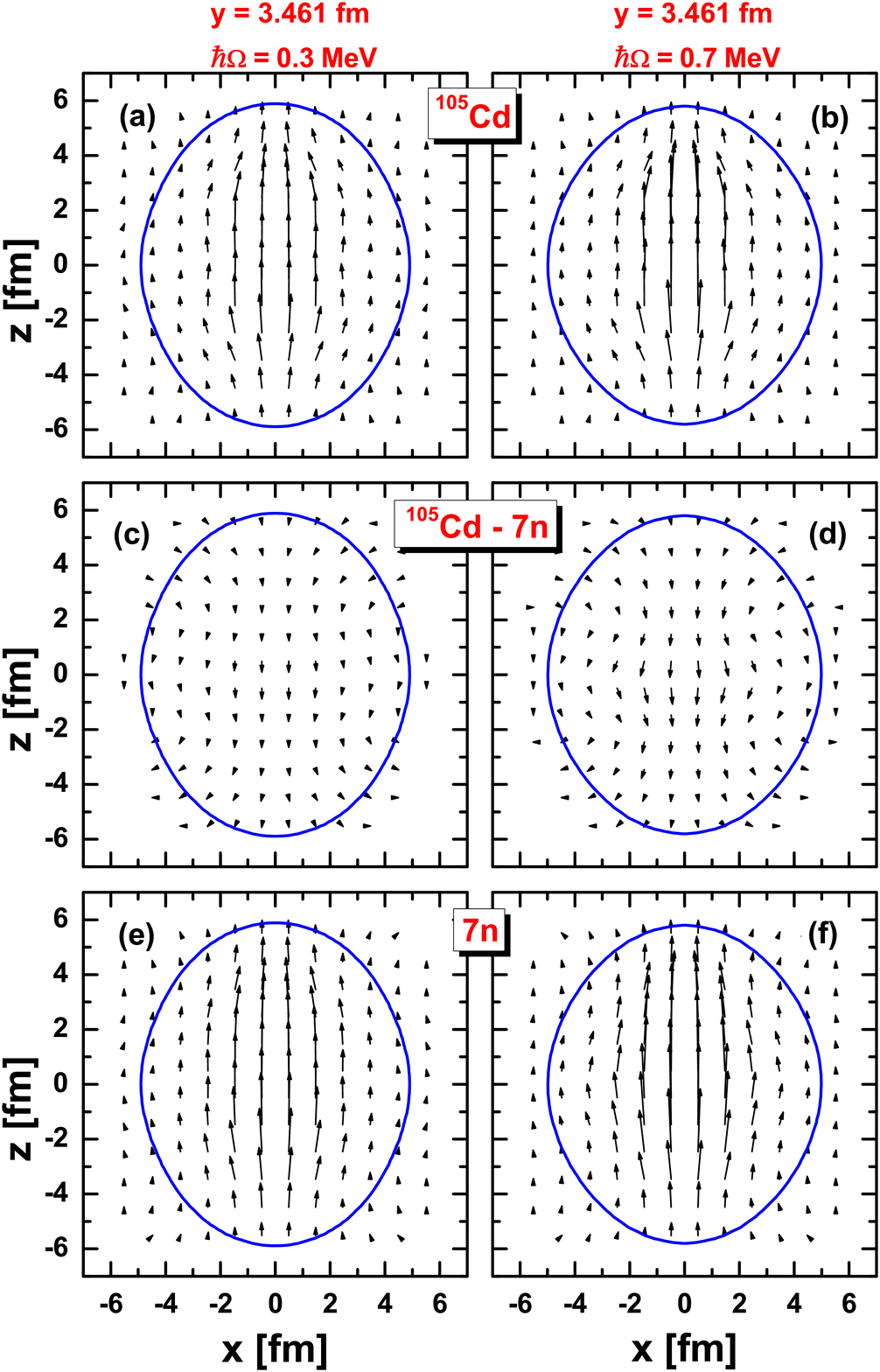}
\caption{(Color online) Same as Fig.~\ref{fig10} but in the $xz$ plane parallel to the rotational axis with $y=3.461~\rm fm$. }
\label{fig11}
\end{figure}

In Fig.~\ref{fig10} we show Dirac currents for the neutrons in the nucleus $^{105}\rm Cd$. We present the currents in the $yz$ plane at $x=1.473~\rm fm$ for two different frequencies, in the left panels for $\hbar\Omega=0.3$~MeV and in the right panels for $\hbar\Omega=0.7$~MeV. The upper two panels (a) and (b) show the total currents of all the neutrons, the
panels (c) and (d) in the middle show the neutron currents in the $N=50$ core and the lower two panels (e) and (f) show the currents of the seven valence nucleons.

\begin{figure}[!htbp]
\includegraphics[width=8cm]{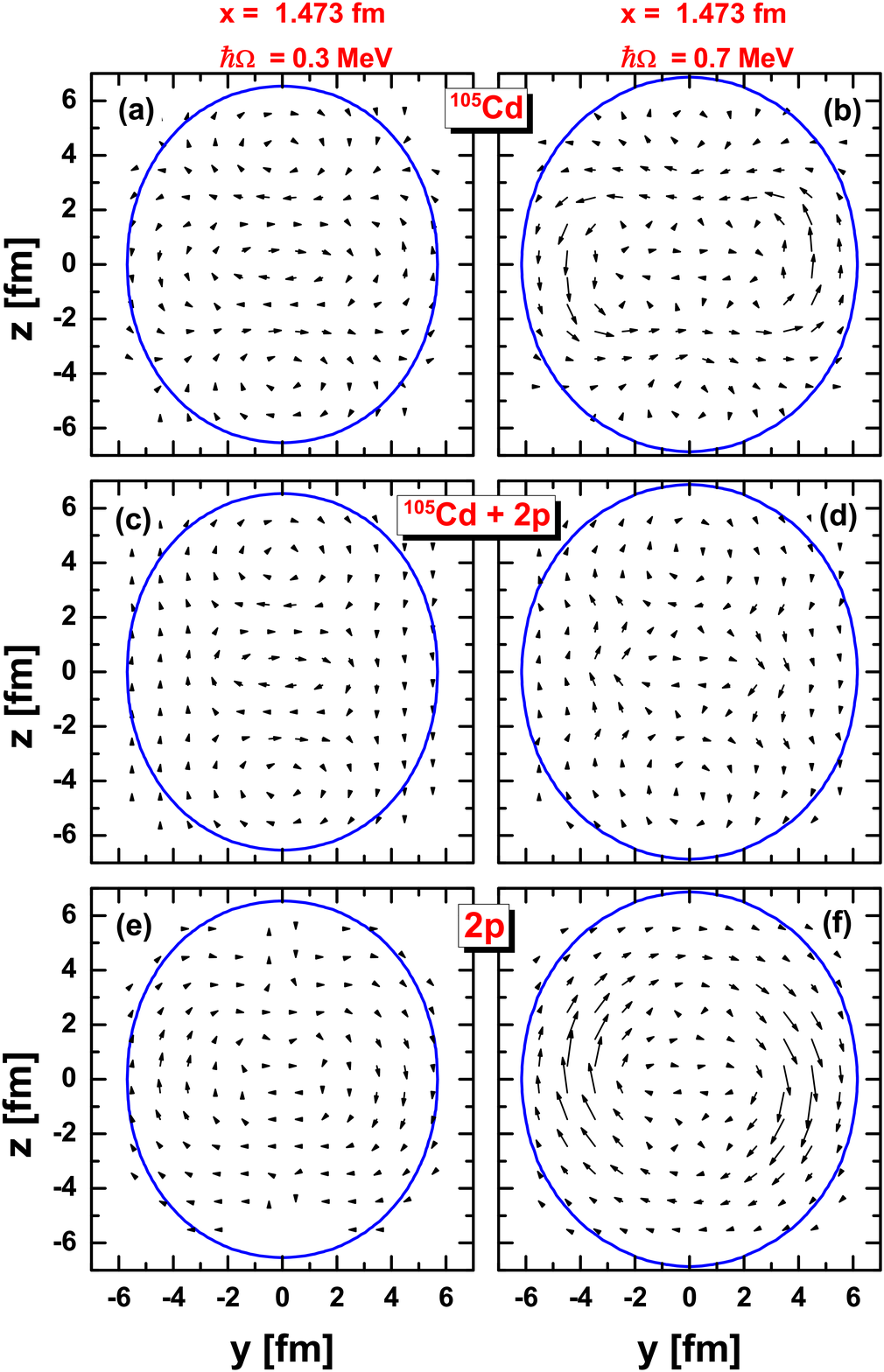}
\caption{(Color online) Dirac currents for $^{105}\rm Cd$ in the $yz$ plane with $x=1.473~\rm fm$ for all protons (top), for the $Z=50$ core (middle), and for the two valence proton holes (bottom) at $\hbar\Omega=0.3~\rm MeV$ (left panels) and $\hbar\Omega=0.7~\rm MeV$ (right panels). The shape and the size of the nucleus $^{105}\rm Cd$ are represented by the solid lines which corresponds to to lines with constant proton density $\rho_p=0.01~{\rm fm}^{-3}$.}
\label{fig12}
\end{figure}

It is clearly seen that the currents in the $N=50$ core are very small at $\hbar\Omega = 0.3~\rm MeV$ and remain that even though the rotational frequency surge from $\hbar\Omega=0.3$~MeV to 0.7~MeV. Of course, the particles in the core feel also a strong Coriolis field $\bm{\Omega J}$, which leads to considerable mixing of the different single particle wave functions. However, the core corresponds approximately  to a closed spherical shell. It is represented by a Slater determinant not depending on the mixing
or the individual wave functions of the orbits. All the time-reversed partners are nearly equally occupied and this leads to a cancellation of the angular momenta and the currents. The net angular momenta and currents stay always negligible. In contrast, in the valence shell there are many empty orbits and the Coriolis field forces the seven valence neutrons to occupy only orbitals with angular momenta aligned in the $x$-direction. Orbits with opposite angular momenta are kept empty. Therefore, as shown in Fig.~\ref{fig8}, net alignments grow considerably with increasing angular velocity and the same holds for the currents. The nuclear magnetic fields induced by these currents enhance this effect in a self-consistent way, because the isoscalar as well as the isovector current-current interactions between nucleons are always attractive. Repulsive current-current interactions between particles and anti-particles play no role because of the no-sea approximation. As a consequence, the total neutron currents are primarily generated by the valence neutrons and, as the nucleus is rotating around the $x$ axis, the total current pattern of Fig.~\ref{fig10} is dominated by a large central symmetric vortex in the $yz$ plane. This current distribution is by no means irrotational.

\begin{figure}[!htbp]
\includegraphics[width=8cm]{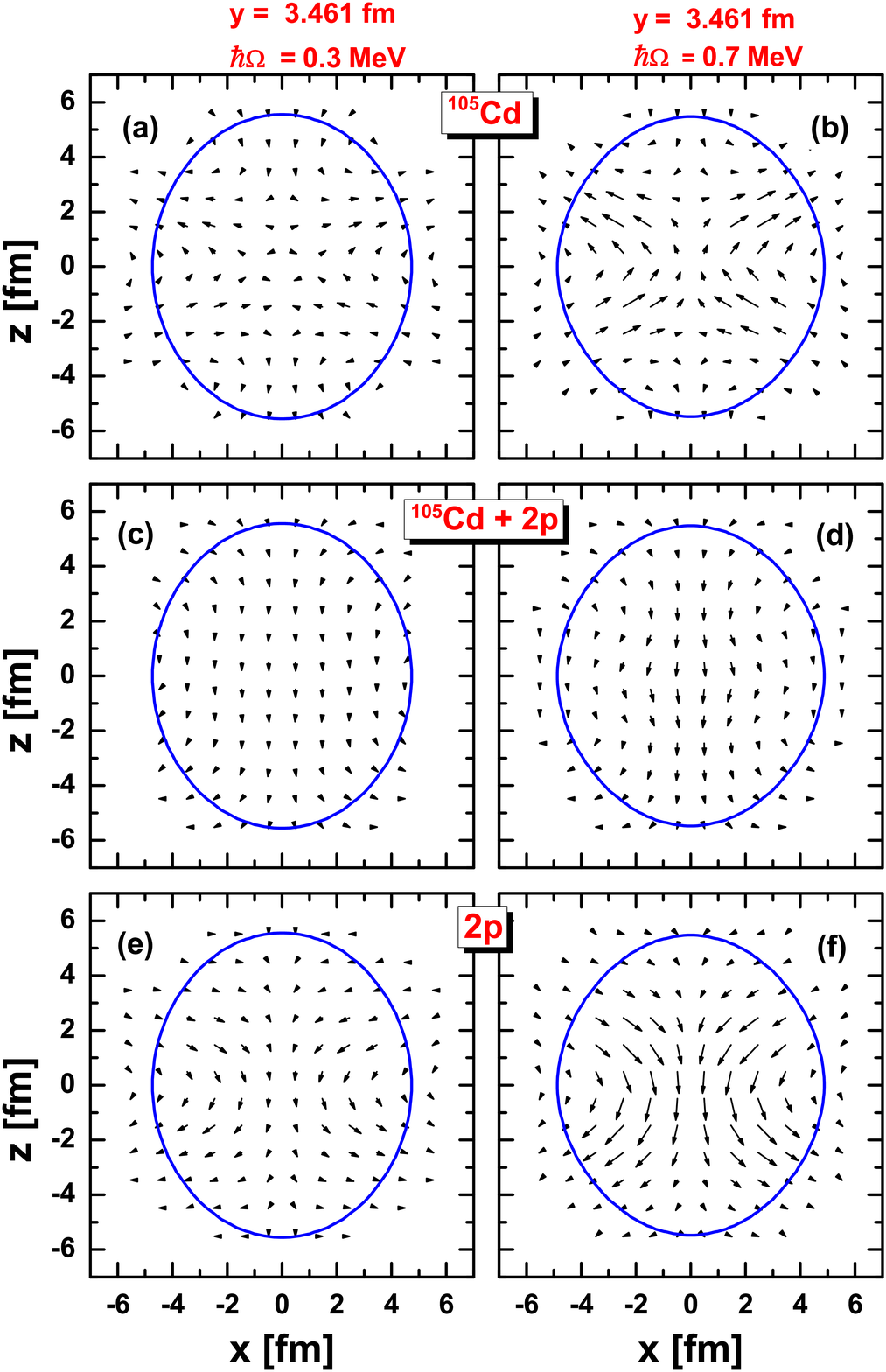}
\caption{(Color online) Same as Fig.~\ref{fig12} but in the $xz$ plane with $y=3.461~\rm fm$.}
\label{fig13}
\end{figure}

In full analogy we show in Fig.~\ref{fig11}, Dirac currents for $^{105}\rm Cd$ in the $xz$ plane parallel to the rotational axis at $y = 3.461$ fm.
Contributions of all neutrons are given at the top, those of the $N = 50$ core in the middle, and those of the seven valence neutrons at the bottom. Two angular velocities $\hbar\Omega=0.3~\rm$ MeV (left panels) and $\hbar\Omega=0.7~\rm$ MeV (right panels) are presented. Again the total currents are dominated by the
valence particles. However, since the nucleus is rotating around the $x$ axis, the current pattern in the $xz$ plane is dominated by a longitude flow
with negligible components in the $x$ direction.

In Fig.~\ref{fig12}, same as in Fig.~\ref{fig10} but for protons,
the Dirac currents for $^{105}\rm Cd$ are given for all protons (top), for the $Z = 50$ core (middle), and for the two valence proton holes (bottom) at $\hbar\Omega=0.3~\rm$ MeV (left panels) and $\hbar\Omega=0.7~\rm$ MeV (right panels).

Since there are only 48 protons in $^{105}\rm Cd$, the total currents are the difference between the contributions from $Z=50$ core and the two valence proton holes. Similar to the currents of the $N=50$ core, the currents for the $Z=50$ core are also negligible at $\hbar\Omega=$0.3~MeV or 0.7~MeV due to the cancellation of the time reversal partners which also provides a negligible angular momentum (see Fig.~\ref{fig8}). As a consequence, the total proton currents are essentially build from the two valence holes which are enhanced with the growing frequency. However, since the total spin of protons is smaller than that of neutrons, the corresponding Coriolis term and the strengths of the proton currents are significantly weaker than the neutron ones at a given rotational frequency.

\begin{figure}[!htbp]
\includegraphics[width=8cm]{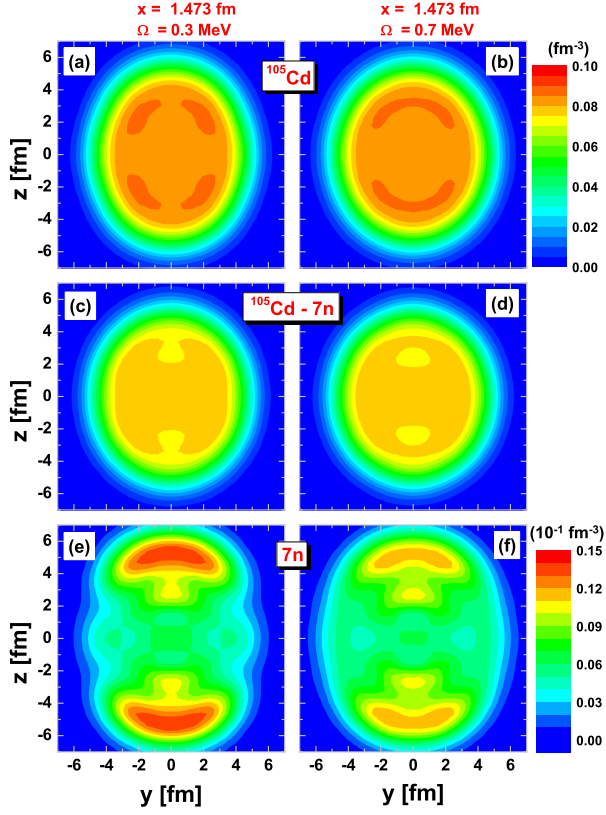}
\caption{(Color online) Density distribution for $^{105}\rm Cd$ in the $yz$ plane with $x=1.473~\rm fm$ for all neutrons (top), $N=50$ core (middle), and the seven valence neutrons (bottom) at $\hbar\Omega=0.3~\rm MeV$ (left panels) and $\hbar\Omega=0.7~\rm MeV$ (right panels). Note that different color scale is used for the valence neutrons for better visualization of densities. }
\label{fig14}
\end{figure}

In Fig.~\ref{fig13}, same as in Fig.~\ref{fig12} but in the $xz$ plane with $y = 3.461$ fm,
the Dirac currents for $^{105}\rm Cd$ are given for all protons (top), for the $Z = 50$ core (middle), and for the two valence proton holes (bottom) at $\hbar\Omega=0.3~\rm$ MeV (left panels) and $\hbar\Omega=0.7~\rm$ MeV (right panels).

\begin{figure}[!htbp]
\includegraphics[width=8cm]{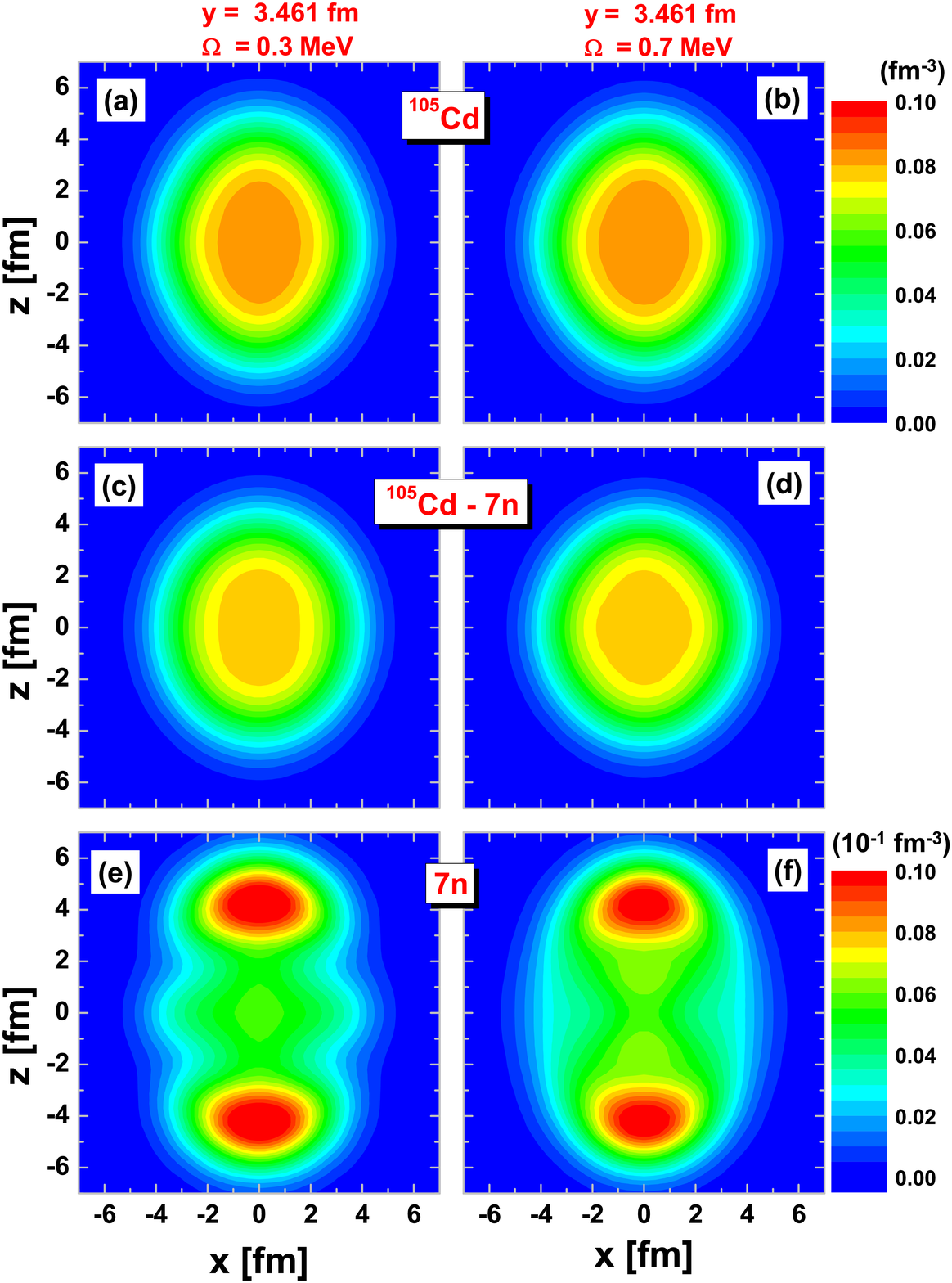}
\caption{(Color online) Same as Fig.~\ref{fig14} but in the $xz$ plane with $y=3.461~\rm fm$. }
\label{fig15}
\end{figure}

For the strengths, the currents in the $xz$ plane increase with rotational frequency, which are similar as the currents in the $yz$ plane. For the current pattern, similar as the neutron cases, the proton current pattern in the $yz$ plane  is dominated by a central symmetric vortex and in the $xz$ plane by a appreciable longitude flow, which is connected with the fact that nucleus here is rotating around the $x$ axis. However, different from the neutron currents in the $xz$ plane, the $x$ components of proton currents are no longer negligible. The difference can be understood from the so-called two-shears-like mechanism in Fig.~\ref{fig7}. The two proton holes in $g_{9/2}$ orbital, which originally orient back to back along the $z$ direction, gradually align towards the $x$ axis and create flow not only in the longitude $z$ direction but also in the $x$ direction.

It is interesting to investigate the evolution of the density distributions with increasing angular velocity and the relation between the currents and the density distributions. In Fig.~\ref{fig14} we show the density distribution of $^{105}\rm Cd$ in the $yz$ plane at $x=1.473~\rm fm$ for all neutrons at $\hbar\Omega=0.3~\rm MeV$ (panel a) and at $\hbar\Omega=0.7~\rm MeV$ (panel b). It is found that the neutron density has a prolate-like shape which is symmetric around the $z$ axis and changes with increasing rotational frequency quantitatively toward to a near-spherical distribution. This is consistent with the deformation evolutions shown in Fig.~\ref{fig6}.

\begin{figure}[!htbp]
\includegraphics[width=8cm]{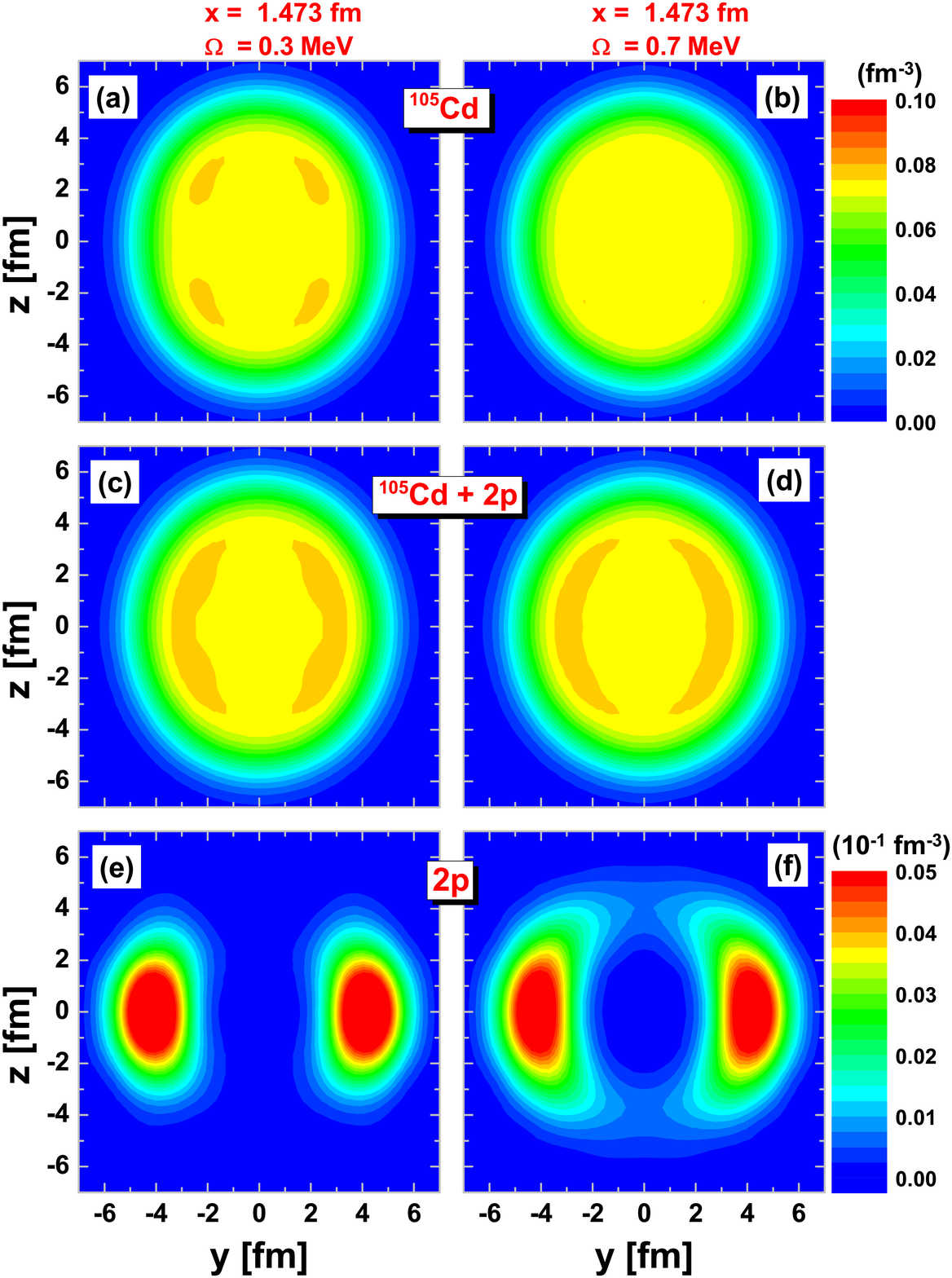}
\caption{(Color online) Density distribution for $^{105}\rm Cd$ in the $yz$ plane with $x=1.473~\rm fm$ for all protons (top), $Z=50$ core (middle), and the two valence proton holes (bottom) at $\hbar\Omega=0.3~\rm MeV$ (left panels) and $\hbar\Omega=0.7~\rm MeV$ (right panels).  Note that different color scale is used for the valence proton holes for better visualization of densities.}
\label{fig16}
\end{figure}

The neutron density is composed of the $N=50$ core as shown in panels (c) and (d) as well as the seven valence neutrons in the panels (e) and (f) in Fig.~\ref{fig14}. It is found that both the densities for the seven valence neutrons and the $N=50$ core have prolate-like distribution with the symmetry axis $z$ axis. Although the densities for the valence neutrons is one order of magnitude smaller than those of the $N=50$ core, the neutron currents in Figs.~\ref{fig10} and \ref{fig11} are mainly produced by the seven valence neutrons rather than the $N=50$ core. Examining the density distributions and the currents of the valence neutrons, it is interesting to note that the minima in density distributions correspond to the maxima of the current strengths.

The same conclusion can also be reached by focusing on the neutron density distributions in the $xz$ plane shown in Fig.~\ref{fig15}, where the density distributions for all neutrons, $N=50$ core, and the seven valence neutrons in $^{105}\rm Cd$ are displayed in the $xz$ plane with $y=3.461~\rm fm$.

\begin{figure}[!htbp]
\includegraphics[width=8cm]{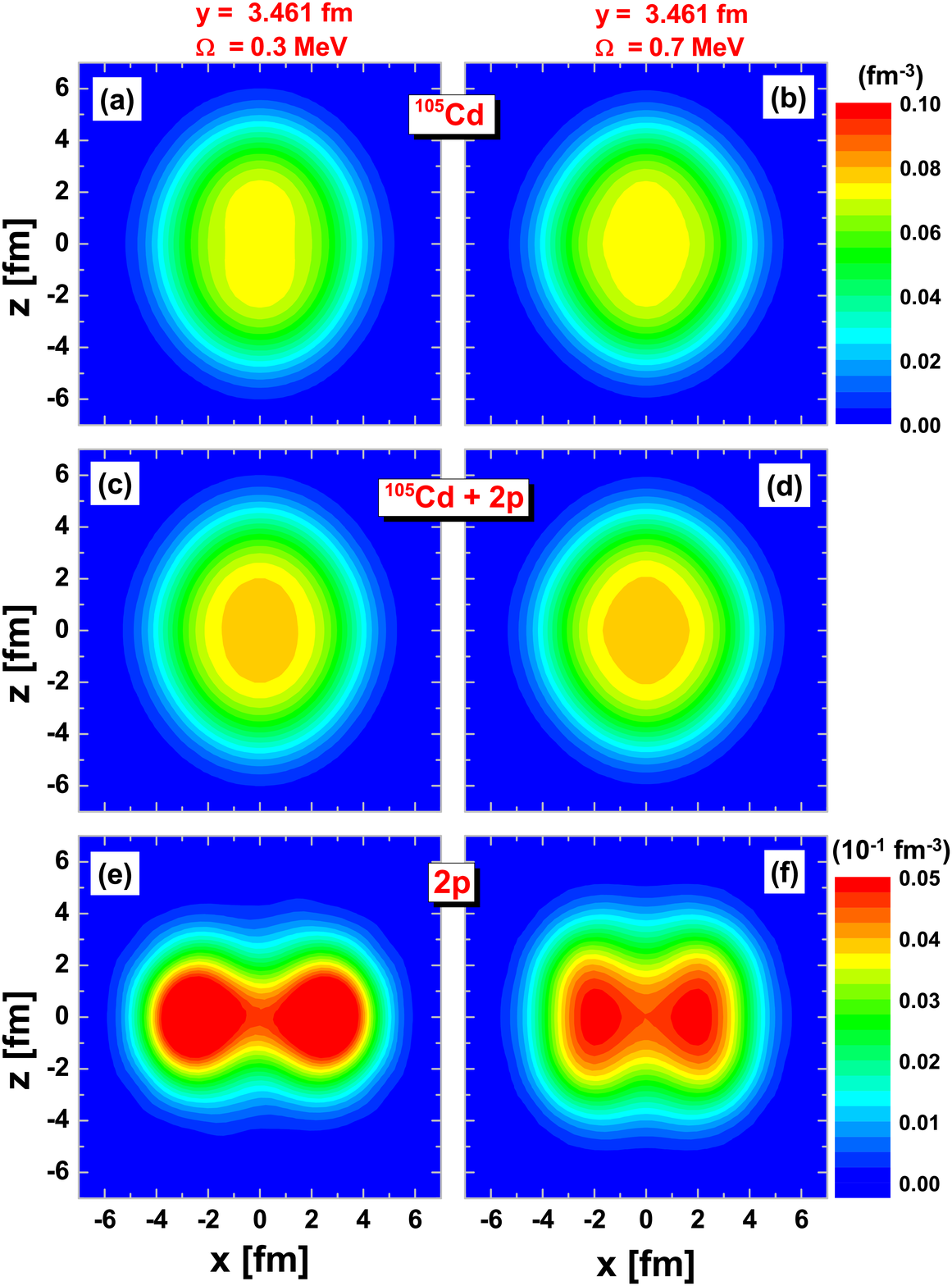}
\caption{(Color online) Same as Fig.~\ref{fig16} but in the $xz$ plane with $y=3.461~\rm fm$. }
\label{fig17}
\end{figure}

In Fig.~\ref{fig16}, same as in Fig.~\ref{fig14} but for protons, the density distributions for $^{105}\rm Cd$ are displayed for all protons (top), for the $Z = 50$ core (middle), and the for the two valence proton holes (bottom) at $\hbar\Omega=0.3~\rm$ MeV (left panels) and $\hbar\Omega=0.7~\rm$ MeV (right panels).

Similar as the neutron case, the proton density exhibits a prolate-like shape with $z$ axis as the symmetry axis and evolutes toward a near-spherical distribution with the rotational frequency. Here the total proton density distribution can be decomposed into the density distributions of the $Z=50$ core in  panels (c) and (d) and the two valence proton holes in panels (e) and (f) in Fig.~\ref{fig16}.
It could be seen that the density distributions for the two valence proton holes have an oblate-like shape in contrast with the prolate shape for the $Z=50$ core.
By subtracting the density of two valence proton holes from the $Z=50$ core, the resulting proton density is driven toward a more prolate shape.

Again the densities for the valence proton holes are one order of magnitude smaller than those of the $Z=50$ core, the proton currents in Figs.~\ref{fig12} and \ref{fig13} are mainly produced by the valence proton holes rather than the $Z=50$ core. Examining the density distributions and the currents of the valence proton holes, it is interesting to note that the maxima in density distributions correspond to the maxima of the current strengths due to the hole character.

The same arguments also apply for the proton density distributions shown in Fig.~\ref{fig17}, where the density distributions of all protons, $Z=50$ core, and the two valence proton holes in $^{105}\rm Cd$ are displayed in the $xz$ plane with $y=3.461~\rm fm$.

\section{Summary}

In summary, covariant density functional theory is used for a systematic investigation and a detailed analysis of the AMR band in the nucleus $^{105}\rm Cd$. Following a short report~\cite{Zhao2011Phys.Rev.Lett.122501} these calculations within the  framework of the TAC model present the first fully microscopic and self-consistent description of a AMR band in nuclei.

Using the point-coupling density functional PC-PK1, the configuration for the observed AMR band in $^{105}\rm Cd$ is fixed by analyzing the single-particle Routhians. With the configuration thus obtained, the tilt angle $\theta_\Omega$ for a given rotational frequency is determined self-consistently by minimizing the total Routhian with respect to the angle $\theta_\Omega$. It demonstrates in an unambiguous way that the rotational axis is always along the $x$ axis.

The energy spectrum, the total angular momentum, the kinetic and dynamic moments of inertia, and the $B(E2)$ values of the AMR band in $^{105}\rm Cd$ are calculated and agreement is found with available experimental data. The change in the energy along the spectrum in the band comes to a large part from the kinetic energy but with a moderate competition from the interaction energy. The observed increase of the ${\mathfrak{J}}^{(2)}/B(E2)$ ratios is well reproduced reflecting the fact that the moment of inertia ${\mathfrak{J}}^{(2)}$ is essentially constant whereas the $B(E2)$ values rapidly decline along the band. This fact clearly distinguishes AMR bands from smoothly terminating bands.

In contrast to the phenomenological model in Ref.~\cite{Clark2000Annu.Rev.Nucl.Part.Sci.1} where the nucleus is divided artificially into a core and serval valence particles or holes, in the present microscopic calculation both the energy and the angular momentum along a band come entirely from individual nucleons. By investigating microscopically the contributions of neutron and protons to the total angular momentum, the two-shears-like mechanism in the AMR band is clearly illustrated.

Finally, the Dirac currents are presented and discussed in details for the AMR band in $^{105}\rm Cd$. They are originally induced by the Coriolis operator, but considerably increased by the nuclear magnetic fields resulting from a self-consistent polarization mechanism. By decomposing the currents into the contributions from the core and from the valence particles or holes, it is found that the core stays mostly inert. The structure of the total currents is essentially determined by the valence particles or holes. Their size and their spatial pattern depends on the specific single-particle orbitals and, of course, on the rotational frequency. In addition we discuss the correlations between the currents and the density distributions in the present calculations.

\begin{acknowledgments}
This work was supported in part by the Major State 973 Program of China (Grant No. 2007CB815000), the Natural Science Foundation of China (Grant Nos. 10975007, 10975008, 11005069, 11175002, 11105006), the Research Fund for the Doctoral Program of Higher Education under Grant No. 20110001110087, the China Postdoctoral Science Foundation (Grant No. 20100480149, 201104031), the Fundamental Research Funds for the Central University, and the Oversea Distinguished Professor Project from Ministry of Education No. MS2010BJDX001. We also acknowledge partial support from the DFG Cluster of Excellence ``Origin and Structure of the Universe''(www.universe-cluster.de).
\end{acknowledgments}

\newpage %Just because of unusual number of tables stacked at end

\end{document}